\documentclass[aps,pre,reprint]{revtex4-1}

\pdfoutput=1

\usepackage{mathtools}
\usepackage{amssymb}
\usepackage[italicdiff]{physics}

\usepackage{xcolor}
\usepackage[colorlinks, citecolor={blue!70!black}, urlcolor={blue!70!black}, linkcolor={red!70!black},hyperindex,breaklinks]{hyperref}
\usepackage{enumerate}
\usepackage{enumitem}

\usepackage{graphicx}
\usepackage{float}

\newcommand{\eq}[1]{ \begin{equation} #1 \end{equation} }
\newcommand{\eqal}[1]{ \begin{align}#1\end{align} }
\newcommand{\eqsub}[1]{ \begin{subequations}#1\end{subequations} }

\newcommand{\eqml}[1]{ \begin{multline}#1\end{multline} }
\newcommand{\mat}[1]{ \begin{pmatrix}#1\end{pmatrix} }

\renewcommand{\hbar}{\mathchar'26\mkern-9mu h}
\newcommand{\wtl}{\widetilde}
\newcommand{\ct}[1]{{#1}^{\dagger}}
\renewcommand{\comm}[2]{\big[ #1, \, #2 \big]}
\renewcommand{\ev}[1]{\langle #1 \rangle}
\newcommand{\cev}[1]{\langle\!\langle #1 \rangle\!\rangle}

\begin{document}

\title{Relaxation to gaussian and generalized Gibbs states in systems of particles with quadratic hamiltonians}

\author{Chaitanya Murthy}
\email{cm@physics.ucsb.edu}
\author{Mark Srednicki}
\email{mark@physics.ucsb.edu}
\affiliation{Department of Physics, University of California, Santa Barbara, CA 93106}

\date{\today}

\begin{abstract}
We present an elementary, general, and semi-quantitative description of relaxation to gaussian and generalized Gibbs states in lattice models of fermions or bosons with quadratic hamiltonians. Our arguments apply to arbitrary initial states that satisfy a mild condition on clustering of correlations. We also show that similar arguments can be used to understand relaxation (or its absence) in systems with time-dependent quadratic hamiltonians, and provide a semi-quantitative description of relaxation in quadratic periodically driven (Floquet) systems. \\
\end{abstract}

\maketitle


\section{Introduction}

In recent years, there has been much work on understanding the nature of the equilibrium state, and the dynamics of the relaxation to this state, in quantum many-body systems with an extensive number of local conservation laws. The motivation for the study of these integrable models is twofold. First, beautiful cold-atom experiments have successfully realized many such models and studied their nonequilibrium dynamics (for reviews, see \cite{Polkovnikov2011,Langen2015,DAlessio2016,Langen2016}). Second, integrable models are much easier to analyze theoretically than their non-integrable brethren (for reviews, see \cite{Essler2016,Vidmar2016}).

When prepared in a generic initial state, an integrable system does not thermalize in the usual sense of the word, due to the extensive number of local conserved quantities. Instead, the stationary behavior at late times can be described by an appropriate \emph{generalized Gibbs ensemble} (GGE). For our purposes, the ``GGE conjecture'' put forth in Ref.~\cite{Rigol2007} and subsequently refined by many authors (\cite{DAlessio2016} and references therein), asserts the following: assuming that local observables in an integrable many-body system relax to stationary values, these values may be computed using the density matrix
\eq{
\hat{\rho}_{\text{GGE}} = \frac{1}{Z_{\text{GGE}}} \exp( - \sum_m \lambda_m \hat{I}_m ) ,
}
where $\{ \hat{I}_m \}$ is the set of all local conserved quantities (here locality means that each $\hat{I}_m$ is a sum of local densities), $Z_{\text{GGE}} = \text{Tr}(e^{-\sum_m \lambda_m \hat{I}_m})$ is the partition function needed for normalization, and $\{ \lambda_m \}$ are Lagrange multipliers chosen so as to satisfy the constraints $\text{Tr}( \hat{I}_m \, \hat{\rho}_{\text{GGE}} ) = \ev{\hat{I}_m}(t=0)$. The density matrix $\hat{\rho}_{\text{GGE}}$ is readily obtained by the general prescription \cite{Jaynes1957} of maximizing the entropy $S = - \Tr(\hat{\rho} \log{\hat{\rho}})$ subject to these constraints.

In this paper we consider the simplest class of integrable models: those whose hamiltonians can be expressed as quadratic forms in a set of canonical particle creation and annihilation operators. Such so-called \emph{non-interacting integrable} models describe not only truly non-interacting particles, but also mean-field approximations to models of interacting particles. In view of this, we will refer to the models of interest as ``quadratic'' rather than ``non-interacting integrable''. In one dimension, certain other integrable models---some spin chains and systems of hard-core particles---can be mapped to quadratic ones. However, simple observables in the original model often map to complicated operators in the quadratic model, and one must take this into account when using the mapping to study relaxation \cite{Wright2014}.

In quadratic models, the local conserved charges $\hat{I}_m$ are also typically quadratic (we demonstrate this explicitly in Section~\ref{sec:conserved_quantities}), and therefore the GGE density operator $\hat{\rho}_{\text{GGE}}$ is \emph{gaussian}. It is thus quite natural to divide relaxation in quadratic models into two process: (i) relaxation of the initial state to a gaussian one, and (ii) relaxation of the gaussian state to the appropriate GGE. Our arguments will make clear that these two processes occur for fundamentally distinct physical reasons, and also that the first process often (but not always) occurs faster than the second.
We will also show that similar arguments can be used to understand relaxation in systems with quadratic time-periodic hamiltonians. Considering the great recent interest in driven quantum systems (\cite{Sieberer2016,Eckardt2017} and references therein), we believe that this is a useful synthesis of ideas.

Recently, Gluza et al.~\cite{Gluza2016} have argued that the first process mentioned above (which they term ``gaussification'') can be understood as a consequence of exponential clustering of correlations in the initial state, together with ``delocalizing transport'' (the sufficiently rapid suppression of the component of an operator on any given site, due to its ``spreading out'' over a large area), and have rigorously proven this implication for fermionic lattice systems with quadratic time-independent hamiltonians. The importance of clustering of correlations in the initial state for relaxation had earlier been emphasized by Cramer and Eisert \cite{Cramer2010}, and by Sotiriadis and Calabrese \cite{Sotiriadis2014}, who had shown that it is in fact a necessary and sufficient condition for relaxation to GGE in a broad class of translation-invariant quadratic models.
Other early studies of the validity of the GGE in quadratic systems (e.g.~\cite{Barthel2008,Cazalilla2012}) typically assumed gaussian initial states, and hence addressed only the second process described in the previous paragraph.
Our results are consistent with, and generalize, the results of these earlier works.

Note that, in the decade since Ref.~\cite{Rigol2007} appeared, the validity of the GGE has been the subject of a large number of theoretical and numerical investigations, of which the works cited above comprise only a handful.
It is beyond the scope of this paper to give a reasonably complete summary and/or critical discussion of all these efforts; for this, we refer the reader to one of the recent reviews of the field, e.g.~\cite{DAlessio2016,Essler2016,Vidmar2016}.

Here, we distill the basic ideas present in \cite{Sotiriadis2014} and especially in \cite{Gluza2016}, and give an elementary, general, and semi-quantitative description of relaxation in quadratic lattice models.
To state our results, we need to introduce some notation.
Let $\hat{\psi}^a_x$ denote the particle creation ($a = +$) or annihilation ($a = -$) operator for the site at position $x$.
Under Heisenberg time-evolution with any time-dependent quadratic hamiltonian $\hat{H}(t)$, these operators evolve linearly:
\eq{
\hat{\psi}^a_x(t) = \sum_{b=\pm} \sum_y G_{xy}^{ab}(t) \hat{\psi}^b_y .
}
This equation defines the \emph{single-particle propagator} $G_{xy}^{ab}(t)$, which plays a central role in the following.

For most of this work, we assume \textbf{delocalizing dynamics:}
\eq{\label{eq:intro_deloc_dyn}
\abs{G_{xy}^{ab}(t)} \to 0 \quad \ \text{as} \quad t \to \infty .
}
The terminology is adapted from Ref.~\cite{Gluza2016}.
Equation~(\ref{eq:intro_deloc_dyn}) may be interpreted as saying that a particle (or hole) created at $x$ has vanishing probability amplitude to be found at any given $y$ after infinite time, because its wavefunction ``spreads out'' indefinitely.
In the case of bosons, we also require the quasiparticle spectrum of $\hat{H}(t)$ to be uniformly bounded and positive-definite.
Note that we \emph{do not} restrict the hamiltonian to be time-independent in general.

We treat \emph{arbitrary initial states} that satisfy a mild condition on \textbf{(algebraic) clustering of correlations:}
\eqal{\label{eq:intro_cluster_decomposition}
\cev{\hat{\psi}^{a_1}_{x_1} \hat{\psi}^{a_2}_{x_2} \cdots \hat{\psi}^{a_n}_{x_n} } 
= \ &o(\abs{x_i - x_j}^{-(d+\epsilon)}) \nonumber \\*
&\text{as} \ \ \abs{x_i - x_j} \to \infty ,
}
for some $\epsilon > 0$.
Here, $\cev{\cdots}$ denotes the \emph{connected correlation function} in the initial state, and $d$ is the spatial dimension.
We emphasize that the initial state is not assumed to have any relation whatsoever to the hamiltonian under which the system subsequently evolves. In particular, the initial state may be strongly interacting.
We also emphasize that the cluster decomposition property (\ref{eq:intro_cluster_decomposition}) is an \emph{extremely weak constraint} on the state; one can expect it to hold for most initial states of interest.
In fact, a stronger version of (\ref{eq:intro_cluster_decomposition}) has been \emph{rigorously proven} for large classes of states, including ground states of interacting local hamiltonians with a spectral gap \cite{Nachtergaele2006,Hastings2006}, as well as thermal states of arbitrary short-ranged fermionic lattice models at sufficiently high temperature \cite{Kliesch2014}.

Our emphasis throughout this paper is on simplicity and physical transparency rather than on mathematical rigor; the reader in search of the latter is encouraged to consult Ref.~\cite{Gluza2016} and similar works in parallel with our treatment. Nevertheless, we believe that most of the arguments presented here can serve as sketches for rigorous proofs.

Our main technical results are summarized below.

\vspace{0.8em}
\noindent \textbf{1.~``Gaussification'' under general conditions:}
\ In any lattice system of fermions or bosons prepared in an initial state satisfying (\ref{eq:intro_cluster_decomposition}) and evolving under a quadratic hamiltonian $\hat{H}(t)$ that leads to delocalizing dynamics (\ref{eq:intro_deloc_dyn}), all local ($n>2$)-point connected correlation functions vanish at late times:
\eq{\label{eq:intro_gaussification}
\cev{\hat{\psi}^{a_1}_{x_1}(t_1) \cdots \hat{\psi}^{a_n}_{x_n}(t_n)} \to 0 \quad \ \text{as} \quad t_j \to \infty .
}
Following Ref.~\cite{Gluza2016}, we refer to this vanishing as ``gaussification'', because it is equivalent to the dynamical recovery of Wick's theorem---the distinguishing property of a gaussian state.
This result significantly generalizes that of Ref.~\cite{Gluza2016}, where gaussification was proven for time-independent fermion lattice models and initial states with exponential clustering of correlations.

\vspace{0.8em}
\noindent \textbf{2.~Universal power-law gaussification:}
\ In many cases of interest one can identify, for each $x$, $a$, and $t$, a definite volume $\mathcal{V}_x^a(t)$ of $y$-space in which the single-particle propagator $G_{xy}^{ab}(t)$ is meaningfully supported.
In these cases we obtain a more quantitative version of Eq.~(\ref{eq:intro_gaussification}):
\eq{\label{eq:intro_cev_decay}
\cev{\hat{\psi}^{a_1}_{x_1}(t_1) \cdots \hat{\psi}^{a_n}_{x_n}(t_n)} \sim [\mathcal{V}(t)]^{-(n/2-1)} \ \ \text{as} \ \ \, t \to \infty ,
}
where $t \coloneqq (t_1 + \cdots + t_n)/n$. This asymptotic result holds in the limit $\abs{t_i - t_j} \ll t$, with $\mathcal{V}(t) \colonapprox \mathcal{V}_{x_i}^{a_i}(t_i) \approx \mathcal{V}_{x_j}^{a_j}(t_j)$.
Typically, $\mathcal{V}(t)$ grows like a power of $t$.
Thus, Eq.~(\ref{eq:intro_cev_decay}) gives power-law decay of the connected correlation functions in time, with exponents that depend only on $n$ and on how fast the single-particle propagator of the system spreads.

\vspace{0.8em}
\noindent \textbf{3.~Gaussianity of GGE:}
\ We prove that all local conserved quantities of a time-independent quadratic hamiltonian with delocalizing dynamics (\ref{eq:intro_deloc_dyn}) are also quadratic, and therefore that the associated GGE density matrix is indeed gaussian. In past work this property appears simply to have been taken for granted.

\vspace{0.8em}
\noindent \textbf{4.~Relaxation to GGE:}
\ For any time-independent hamiltonian that satisfies (\ref{eq:intro_deloc_dyn}), and any initial state that satisfies (\ref{eq:intro_cluster_decomposition}) and an additional assumption to be described below, we show that the system relaxes to the GGE; for any local operator $\hat{\mathcal{O}}$,
\eq{
\ev{\hat{\mathcal{O}}(t)} \to \ev{\hat{\mathcal{O}}}_{\text{GGE}} \quad \text{as} \quad t \to \infty .
}
This result is consistent with and generalizes existing proofs of relaxation to the GGE to a larger class of models and initial states.
The additional assumption is formulated precisely in Section~\ref{sec:2pt_relaxation} (Eq.~(\ref{eq:gen_homogeneous})). Roughly speaking, it excludes situations in which the initial profiles of local conserved densities are inhomogeneous on length scales comparable to the system size. In such cases, the GGE conjecture fails for a trivial reason: it takes infinitely long for the locally conserved density to flow across the whole system in the thermodynamic limit.

\vspace{0.8em}
\noindent \textbf{5.~Universal power-law relaxation:}
\ Under the conditions of the previous result, we also obtain quantitative estimates for how the local 2-point function relaxes to its GGE value. Consider the instantaneous deviation
\eq{
\delta C_{xy}^{ab}(t) \coloneqq \ev{\hat{\psi}_x^a(t) \hat{\psi}_y^b(t)} - \ev{\hat{\psi}_x^a \hat{\psi}_y^b}_{\text{GGE}} .
}
Assume, as is often the case, that the density of quasiparticle states of $\hat{H}$ near the band edge has the form $g(\varepsilon) \sim \varepsilon^s$. We show that, \emph{generically},
\eq{
\delta C_{xy}^{ab}(t) \sim t^{-\alpha (1+s)} \quad \text{as} \quad t \to \infty ,
}
where $\alpha = 1$ if there is a density wave of one or more of the conserved quantities in the initial state, and $\alpha = 2$ if not. 
In particular, for \emph{translation-invariant} quadratic hamiltonians in $d$ dimensions,
\eq{
\delta C_{xy}^{ab}(t) \sim t^{-\alpha d/2} \quad \text{as} \quad t \to \infty ,
}
where $\alpha$ is defined above.
Note that this result holds for \emph{generic} hamiltonians and initial states of the types considered; different exponents can and do occur if the hamiltonian and/or initial state is fine-tuned.

\vspace{0.8em}
\noindent \textbf{6.~Floquet-GGE:}
\ For any \emph{time-periodic} quadratic hamiltonian $\hat{H}(t) = \hat{H}(t+T)$ that satisfies (\ref{eq:intro_deloc_dyn}), we prove that all local conserved quantities of the associated Floquet hamiltonian $\hat{H}_F$ are quadratic, and hence that the Floquet-GGE \cite{Moessner2017} density matrix, $\hat{\rho}_F$, is gaussian.
For any initial state that satisfies the assumptions of result 4 above, we show that the system eventually relaxes to this (time-periodic) Floquet-GGE.
In the limit $T \to 0$ of fast driving, we expect to observe power-law relaxation to $\hat{\rho}_F$, with the exponents given by results 2 and 5 above applied to $\hat{H}_F$.
In the opposite limit $T \to \infty$ of slow driving, we expect to observe power-law relaxation toward a GGE of the instantaneous hamiltonian $\hat{H}(t)$, followed by much slower exponential relaxation $\sim e^{-t/T}$ toward $\hat{\rho}_F$. 
Our results are consistent with, and generalize, the original treatment of this problem by Lazarides et al.~\cite{Lazarides2014} (in which the initial state was assumed to be gaussian, and no estimates like the ones above were given for the relaxation process itself).

\vspace{0.8em}
\noindent \textbf{7.~Effects/non-effects of localized states:}
\ We find that dynamics generated by a quadratic fermion hamiltonian $\hat{H}$ whose quasiparticle spectrum includes discrete localized levels (so that Eq.~(\ref{eq:intro_deloc_dyn}) is violated) \emph{will still lead to gaussification and equilibration to an appropriate GGE}, as long as (i) the initial state has a finite correlation length $\xi$, and (ii) the spatial distance between any pair of localized levels is large relative to $\xi$.
This result should be viewed as an interesting (but easily understandable; see Section~\ref{sec:localized_states}) exception to the general rule \cite{Ziraldo2012,Ziraldo2013} that the GGE fails if the spectrum of $\hat{H}$ contains a pure-point part coming from localized levels.
If either of the conditions (i) or (ii) above is violated, we recover the expected failure of gaussification and of the GGE.
Thus, our results are fully consistent with those of Ziraldo et al.~\cite{Ziraldo2012,Ziraldo2013}, who considered the case that $\hat{H}$ is disordered; in this case condition (ii) will typically be violated.

\subsubsection*{Organization of the paper}

In Section~\ref{sec:nn_chain} we introduce our arguments in a simple and concrete setting: a 1d tight-binding model of spinless fermions. In Section~\ref{sec:setup} we define the general problem and fix terminology and notation. In Section~\ref{sec:gaussification} we present our argument for gaussification in arbitrary quadratic models (of particles), and predict the exponents of the power law decay in time of all higher-point connected functions. In Section~\ref{sec:equilibration_to_gge} we describe the manner in which, for time-independent hamiltonians, the gaussian state equilibrates to the GGE. In Section~\ref{sec:localized_states} we describe how discrete localized levels in the spectrum of the hamiltonian affect relaxation. In Section~\ref{sec:floquet_gge} we consider quenches to time-periodic quadratic hamiltonians, and describe relaxation to the Floquet-GGE. In Section~\ref{sec:spins} we briefly comment on the application of our arguments to spin models that can be mapped onto quadratic fermion models. Finally, in Section~\ref{sec:conclusions} we summarize our results.

\section{Example: Relaxation in a nearest-neighbor tight-binding chain}\label{sec:nn_chain}

In this section we introduce our arguments by working them out carefully in a simple and concrete example, while emphasizing the key ideas. This will also serve to motivate the subsequent general development.

\subsection{Setup}\label{sec:nn_setup}

Consider a tight-binding model of spinless fermions in $d=1$ dimensions, with nearest-neighbor hopping. The hamiltonian is
\eq{\label{eq:nn_H0}
\hat{H}_0 = - \frac{1}{2} \sum_{x=1}^L \left( \hat{c}^{\dagger}_x \hat{c}_{x+1} + \hat{c}^{\dagger}_{x+1} \hat{c}_x \right) ,
}
where $\hat{c}^{\dagger}_x$ and $\hat{c}_x$ respectively create and annihilate a fermion on the site at position $x$. These operators obey the standard anti-commutation relations
\eqsub{
\eqal{
\hat{c}_x \hat{c}^{\dagger}_y + \hat{c}^{\dagger}_y \hat{c}_x &= \delta_{xy} , \\*
\hat{c}_x \hat{c}_y + \hat{c}_y \hat{c}_x &= 0 .
}
}
We have set the lattice spacing equal to 1 and hopping energy equal to $1/2$. We will also set $\hbar = 1$. Periodic boundary conditions are assumed (site $L+1$ is identified with site $1$). 

Imagine a quench in which the system is prepared in some non-equilibrium initial state, represented by the density operator $\hat{\rho}_0$, at time $t=0$, and subsequently evolved with the hamiltonian $\hat{H}_0$ of Eq.~(\ref{eq:nn_H0}). For the majority of this example (up to and including Section~\ref{sec:nn_gaussification}), we make only two assumptions about $\hat{\rho}_0$. 

The first assumption is very important: $\hat{\rho}_0$ must obey the principle of \emph{cluster decomposition} \cite{Weinberg1995}. Roughly speaking, this principle requires correlations between local operators in the state $\hat{\rho}_0$ to factorize as the operators are taken far apart from one another. We will make this precise in Eq.~(\ref{eq:nn_exp_cluster_decomp}) (we use a stronger version of the principle in this section than we do in our general treatment).  

The second assumption is \emph{not} important, and we impose it only to simplify the example. We assume that the initial state conserves total particle number:
\eq{\label{eq:nn_N_cons}
\comm{\hat{N}}{\hat{\rho}_0} = 0 ,
}
where
\eq{
\hat{N} = \sum_x \hat{c}^{\dagger}_x \hat{c}_x .
}
In the general treatment of Section~\ref{sec:setup} onwards, we make no assumption like Eq.~(\ref{eq:nn_N_cons}).

In the last part of this example, Section~\ref{sec:nn_2pt_relaxation}, we will add a third assumption about $\hat{\rho}_0$, Eq.~(\ref{eq:nn_homogeneous}). Nothing in Sections~\ref{sec:nn_gge} through \ref{sec:nn_gaussification} relies on this extra assumption; it is only needed for the analysis of Section~\ref{sec:nn_2pt_relaxation}. Therefore, we do not state it here.

We will study whether and how local observables of the system relax to their values in an appropriate generalized Gibbs ensemble as time progresses. We first discuss the construction of this GGE density operator.

\subsection{Conserved quantities and GGE density operator}\label{sec:nn_gge}

The hamiltonian (\ref{eq:nn_H0}) can be diagonalized by introducing quasi-momentum mode operators:
\eq{
\hat{c}_x = \frac{1}{\sqrt{L}} \sum_k e^{ikx} \hat{c}(k) ,
}
where $k$ runs over all integer multiples of $(2\pi/L)$ within the Brillouin zone $(-\pi, \pi]$. In terms of these mode operators,
\eq{\label{eq:nn_H0_diag}
\hat{H}_0 = \sum_k \omega(k) \, \hat{c}^{\dagger}\!(k) \hat{c}(k) , \quad \
\omega(k) = -\cos{k} .
}

The various mode occupation number operators
\eq{\label{eq:nn_n(k)}
\hat{n}(k) = \hat{c}^{\dagger}\!(k) \hat{c}(k)
}
clearly commute with $\hat{H}_0$ and with each other. Furthermore, by forming appropriate linear combinations of them, we can define an extensive set of local conserved quantities in involution:
\eqsub{
\label{eq:nn_I1m}
\eqal{
\hat{I}_{2m} &= \sum_k \cos(mk) \hat{n}(k) \\*
&= \frac{1}{2} \sum_{x=1}^L \left( \ct{\hat{c}}_x \hat{c}_{x+m} + \ct{\hat{c}}_{x+m} \hat{c}_x \right) ,
}
}
and
\eqsub{
\label{eq:nn_I2m}
\eqal{
\hat{I}_{2m+1} &= \sum_k \sin(mk) \hat{n}(k) \\*
&= \frac{(-i)}{2} \sum_{x=1}^L \left( \ct{\hat{c}}_x \hat{c}_{x+m} - \ct{\hat{c}}_{x+m} \hat{c}_x \right) .
}
}
where $m=0,1,2,\dots$. These clearly commute with $\hat{H}_0$ and with one another:
\eqsub{
\eqal{
\comm{\hat{H}_0}{\hat{I}_{m}} &= 0 , \\*
\comm{\hat{I}_{m}}{\hat{I}_{m'}} &= 0
}
}
(in fact, $\hat{H}_0 = \hat{I}_{2}$, so the second equation implies the first). They are local because their densities,
\eq{
\propto \left( \ct{\hat{c}}_x \hat{c}_{x+m} \pm \ct{\hat{c}}_{x+m} \hat{c}_x \right) ,
}
act nontrivially only on finite intervals of length $m$.

The set of local conserved quantities $\{ \hat{I}_m \}$ defined in Eqs.~(\ref{eq:nn_I1m}) and (\ref{eq:nn_I2m}) has the further property of being \emph{maximal}: any local conserved quantity $\hat{I}$ that commutes with all of the $\hat{I}_m$ can be expressed as a linear combination of them,
\eq{\label{eq:nn_maximal_def}
\comm{\hat{I}_m}{\hat{I}\,} = 0 \quad \forall \ m \ \implies \
\hat{I} \in \text{span}(\{ \hat{I}_m \}) ,
}
where
\eq{
\text{span}(\{ \hat{I}_m \}) \coloneqq \left\{ \sum_m a_m \hat{I}_m \, \middle| \, a_m \in \mathbb{R} \right\} .
}
This claim is easy to verify if we assume that $\hat{I}$ is a quadratic operator; the only quadratic operators that commute with $\hat{n}(k)$ for all $k$ are indeed of the form $\hat{I} = \sum_k f(k) \hat{n}(k)$ for some function $f$. However, once we drop this assumption, the validity of the claim is much less obvious. One can certainly write down many \emph{nonlocal} conserved quantities that violate Eq.~(\ref{eq:nn_maximal_def})---products of mode occupation numbers, such as $\hat{n}(k) \hat{n}(k')$---and one might wonder whether it is possible to build a local quantity out of linear combinations of these, \`{a} la Eqs.~(\ref{eq:nn_I1m}) or (\ref{eq:nn_I2m}). We will address this concern later in our general treatment: in Section~\ref{sec:conserved_quantities}, we prove that, for a wide class of quadratic hamiltonians (to which $\hat{H}_0$ belongs), all local conserved quantities $\hat{I}$ are themselves quadratic. The claim follows.

Thus, one is tempted to assert that the GGE density operator for the tight-binding chain has the form
\eqsub{
\label{eq:nn_gge_rho}
\eqal{
\hat{\rho}_{\text{GGE}} &= \frac{1}{Z_{\text{GGE}}} \exp( - \sum_m \lambda_m \hat{I}_m ) 
\label{eq:nn_gge_rho_a} \\*
&= \frac{1}{Z_{\text{GGE}}} \exp( - \sum_k \mu(k) \hat{n}(k) ) ,
\label{eq:nn_gge_rho_b} 
}
}
where the Lagrange multipliers $\{ \lambda_m \}$ are fixed by requiring that
\eq{\label{eq:nn_gge_constraint}
\Tr( \hat{I}_m \, \hat{\rho}_{\text{GGE}} ) = \Tr( \hat{I}_m \, \hat{\rho}_0 ) ;
}
this in turn fixes the function $\mu(k)$, which is in general unrelated to the function $\omega(k)$ appearing in $\hat{H}_0$. 

One may also consider \emph{truncated GGEs} in which only the ``most local'' conservation laws are taken into account (i.e.~only $\hat{I}_m$ with $m \leq 2\ell$ are retained in the density matrix) \cite{Fagotti2013}; this is equivalent to truncating the Fourier series of $\mu(k)$ at order $\ell$. More generally, in the limit of infinite system size, $L \to \infty$, one can require that $\lambda_m$ decay in a certain manner as $m \to \infty$; this is equivalent to placing a smoothness condition on $\mu(k)$. Thus, the GGE (truncated or not) can be defined either in terms of the local charges $\hat{I}_m$ or in terms of the mode occupation numbers $\hat{n}(k)$ \cite{Fagotti2013}.

Actually, $\hat{\rho}_{\text{GGE}}$ is not uniquely given by Eq.~(\ref{eq:nn_gge_rho}) for this model. Although the set $\{ \hat{I}_m \}$ defined by Eqs.~(\ref{eq:nn_I1m}) and (\ref{eq:nn_I2m}) is maximal, it is \emph{not complete}: there exist local conserved quantities $\hat{I}'$ that \emph{cannot} be expressed as linear combinations of the $\hat{I}_m$. A simple example \cite{Essler2016} of such a quantity is
\eq{
\hat{I}' = \sum_x (-1)^x \left( \hat{c}_x \hat{c}_{x+1} + \ct{\hat{c}}_{x+1} \ct{\hat{c}}_x \right) .
}
In $k$-space, $\hat{I}'$ takes the form
\eq{
\hat{I}' = \sum_k e^{-ik} \, \hat{c}(\pi - k) \hat{c}(k) + \text{h.c.} .
}
This quantity is conserved because the mode spectrum $\omega(k) = -\cos{k}$ of the hamiltonian $\hat{H}_0$ (Eq.~(\ref{eq:nn_H0_diag})) has the symmetry
\eq{\label{eq:nn_omega_symmetry}
\omega(k) = - \omega(\pi - k) .
}
One can verify that $\hat{I}'$ does not commute with the $\hat{I}_m$, so its existence does not contradict maximality of $\{ \hat{I}_m \}$.

Note that the symmetry (\ref{eq:nn_omega_symmetry}) is actually a \emph{degeneracy} of the spectrum $\abs{\omega(k)} = \abs{\cos{k}}$ of positive-energy quasiparticles of $\hat{H}_0$. In general, the existence of ``extra''  local conserved quantities such as $\hat{I}'$---and the associated ambiguity in the definition of the GGE---is related to degeneracies in the quasiparticle spectrum of the hamiltonian \cite{Fagotti2014}. We discuss the general relationship in Section~\ref{sec:gge_rho}. 

One way to deal with an incomplete maximal set $\{ \hat{I}_m \}$ is to simply \emph{complete} it by adding to $\{ \hat{I}_m \}$ additional local conserved quantities, such as $\hat{I}'$. This is the approach advocated by Fagotti \cite{Fagotti2014}, who studied this problem in significant detail. The operators comprising the expanded set will no longer be in involution, but one can still assign to each one a Lagrange multiplier and define $\hat{\rho}_{\text{GGE}}$ by maximizing the entropy subject to all constraints. We obtain an expression identical to Eq.~(\ref{eq:nn_gge_rho_a}), but where the index $m$ ranges over the complete set. This maneuver is valid because local conserved quantities satisfy a closed algebra \cite{Fagotti2014} (of which the various maximal sets are maximal abelian subalgebras). The advantage of this approach is that the resulting $\hat{\rho}_{\text{GGE}}$ depends on the initial state only through the Lagrange multipliers $\{ \lambda_m \}$. The primary disadvantage is that one can no longer write $\hat{\rho}_{\text{GGE}}$ in terms of a single set of mode occupation numbers, as in Eq.~(\ref{eq:nn_gge_rho_b}).

Our approach to this problem, which we describe in Section~\ref{sec:gge_rho}, is slightly different. In short, we retain $\hat{\rho}_{\text{GGE}}$ in the original form (\ref{eq:nn_gge_rho}), but allow the maximal commuting set $\{ \hat{I}_m \}$, or equivalently the set of mode occupation numbers $\{ \hat{n}_{\alpha} \}$, to depend on the initial state. In this approach, $\hat{\rho}_{\text{GGE}}$ always has a mode number representation similar to Eq.~(\ref{eq:nn_gge_rho_b}); however, different classes of initial states lead to \emph{inequivalent} GGEs.

For now, we can ignore these subtleties, because we assumed that the initial state $\hat{\rho}_0$ conserves total particle number (Eq.~(\ref{eq:nn_N_cons})). For this class of initial states, the GGE is correctly given by Eqs.~(\ref{eq:nn_gge_rho}) and (\ref{eq:nn_gge_constraint}), with $\hat{I}_m$ defined in Eqs.~(\ref{eq:nn_I1m}) and (\ref{eq:nn_I2m}), and $\hat{n}(k)$ in Eq.~(\ref{eq:nn_n(k)}). We leave the proof of this assertion as an exercise for the reader.

\subsection{Relaxation of local observables: preliminaries}\label{sec:nn_relaxation}

Having defined the GGE, we turn to the relaxation of local observables. It is convenient to work in the Heisenberg picture. The operators representing observables evolve according to
\eq{
\hat{\mathcal{O}}(t) = e^{i \hat{H}_0 t} \hat{\mathcal{O}} e^{-i \hat{H}_0 t} ,
}
while the density operator is always $\hat{\rho}_0$. The expectation value of an observable at time $t$ is
\eq{
\ev{\hat{\mathcal{O}}(t)} \coloneqq \Tr( \hat{\mathcal{O}}(t) \, \hat{\rho}_0 ) .
}
By a \emph{local observable} we mean any bosonic hermitian operator $\hat{\mathcal{O}}$ that acts nontrivially only on a finite interval (at time $t = 0$). Consider the quantity
\eq{
R_{\mathcal{O}}(t) \coloneqq \lim_{L \to \infty} \left[ \ev{\hat{\mathcal{O}}(t)} - \ev{\hat{\mathcal{O}}}_{\text{GGE}} \right] ,
}
where
\eq{
\ev{\hat{\mathcal{O}}}_{\text{GGE}} \coloneqq \text{Tr}( \hat{\mathcal{O}} \, \hat{\rho}_{\text{GGE}} ) .
}
We say that the system \emph{relaxes (locally) to the GGE} if
\eq{
R_{\mathcal{O}}(t) \to 0 \quad \text{as} \quad
t \to \infty
}
for every local observable $\hat{\mathcal{O}}$.

Now, any number-conserving local observable has a unique expansion of the form
\eqal{
\hat{\mathcal{O}} = \ &\mathcal{O}^{(0)} 
+ \sum_{x,y} \mathcal{O}^{(1)}_{xy} \ct{\hat{c}}_x  \hat{c}_y \nonumber \\*
&+ \! \sum_{x,x',y,y'} \! \mathcal{O}^{(2)}_{xx'yy'} \ct{\hat{c}}_x \ct{\hat{c}}_{x'} \hat{c}_y \hat{c}_{y'}
\, + \cdots ,
}
where locality implies that all sums over positions are restricted to a finite interval, and therefore that the expansion terminates at a finite order (because the space of operators supported on a finite interval in a system of fermions is finite dimensional). Our simplifying assumption (\ref{eq:nn_N_cons}) on the initial state means that we do not need to consider non-number-conserving observables; their expectation values vanish identically.

Thus, it is sufficient to study the relaxation of local \emph{static $2n$-point correlation functions}:
\eq{
\ev{\ct{\hat{c}}_{x_1}\!(t) \ct{\hat{c}}_{x_2}\!(t) \cdots \ct{\hat{c}}_{x_n}\!(t) \hat{c}_{y_1}\!(t) \hat{c}_{y_2}\!(t) \cdots \hat{c}_{y_n}\!(t)} .
}
More generally, one might also consider \emph{dynamic} correlation functions, in which the various $t$'s are allowed to be different. These describe, for instance, the response of the system to an external probe. 

For systems with a Lieb-Robinson bound \cite{Lieb1972}, there is a general result \cite{Essler2012} which states that, if the system relaxes to a stationary state $\hat{\rho}_{\text{stat}}$ as $t \to \infty$ (as measured by local static correlations), then all local dynamic correlations are also described by $\hat{\rho}_{\text{stat}}$ as $t \to \infty$. In this section, we will simply appeal to this result and concentrate on static correlations. But in fact, most of our arguments apply equally well to dynamic correlations, and we will work directly with the latter in the general treatment from Section~\ref{sec:setup} onwards. We will do this, despite the result of Ref.~\cite{Essler2012}, for two reasons: firstly, to keep our arguments self-contained, and secondly, because we are interested not just in the limiting behavior of quantities as $t \to \infty$, but also in the manner in which they relax to those limits.

Note that the density operator $\hat{\rho}_{\text{GGE}}$ is gaussian---it is the exponential of a quadratic form in the creation and annihilation operators. Therefore, all correlation functions computed with respect to $\hat{\rho}_{\text{GGE}}$ \emph{Wick factorize} into products of two-point functions, and are determined entirely by the latter:
\eqml{
\ev{\ct{\hat{c}}_{x_1} \cdots \, \ct{\hat{c}}_{x_n} \hat{c}_{y_1} \cdots \, \hat{c}_{y_n}}_{\text{GGE}} \vphantom{\Big|} \\*
= \sum_P \text{sgn}(P) \, \ev{\ct{\hat{c}}_{x_1} \hat{c}_{y_{P(n)}}}_{\text{GGE}} \cdots
\ev{\ct{\hat{c}}_{x_n} \hat{c}_{y_{P(1)}}}_{\text{GGE}} ,
}
where $\text{sgn}(P)$ is the sign of the permutation
\eq{
P: \, (1,2,\dots,n) \mapsto (P(1), P(2), \dots, P(n)) .
}
For instance,
\eqal{
\ev{\ct{\hat{c}}_{x_1} \ct{\hat{c}}_{x_2} \hat{c}_{x_3} \hat{c}_{x_4}}_{\text{GGE}}
= \ &\ev{\ct{\hat{c}}_{x_1} \hat{c}_{x_4}}_{\text{GGE}} \, \ev{\ct{\hat{c}}_{x_2} \hat{c}_{x_3}}_{\text{GGE}} \nonumber \\*
&- \ev{\ct{\hat{c}}_{x_1} \hat{c}_{x_3}}_{\text{GGE}} \, \ev{\ct{\hat{c}}_{x_2} \hat{c}_{x_4}}_{\text{GGE}} .
}

Since the initial state $\hat{\rho}_0$ need not be gaussian, the real correlation functions certainly need not behave in this manner at early times. To show relaxation to the GGE, we therefore need to show that, as time progresses, (i) Wick factorization is recovered, and (ii) two-point correlation functions approach their stationary GGE values.

In any state, such as $\hat{\rho}_0$, one can also define the \emph{connected} $2n$-point correlation function. Roughly speaking, this is the part of the $2n$-point correlation function that \emph{fails} to factorize into lower-point correlation functions. For instance,
\eq{
\cev{\ct{\hat{c}}_{x_1} \ct{\hat{c}}_{x_2}} 
= \ev{\ct{\hat{c}}_{x_1} \ct{\hat{c}}_{x_2}} ,
}
and
\eqal{\label{eq:nn_4pt_conn_def}
\cev{\ct{\hat{c}}_{x_1} \ct{\hat{c}}_{x_2} \hat{c}_{x_3} \hat{c}_{x_4}}
= \ &\ev{\ct{\hat{c}}_{x_1} \ct{\hat{c}}_{x_2} \hat{c}_{x_3} \hat{c}_{x_4}} \nonumber \\*
&- \ev{\ct{\hat{c}}_{x_1} \hat{c}_{x_4}} \, \ev{\ct{\hat{c}}_{x_2} \hat{c}_{x_3}} \nonumber \\*
&+ \ev{\ct{\hat{c}}_{x_1} \hat{c}_{x_3}} \, \ev{\ct{\hat{c}}_{x_2} \hat{c}_{x_4}} .
}
The general definition of connected functions is reviewed in Appendix~\ref{app:cumulants}. The vanishing of all connected $(2n > 2)$-point correlation functions is equivalent to Wick factorization, as is evident from the formulae above:
\eqml{
\Big( \ev{\ct{\hat{c}}_{x_1} \cdots \ct{\hat{c}}_{x_n} \hat{c}_{y_1} \cdots \hat{c}_{y_n}} \ \
\text{Wick factorizes} \ \forall \ n \Big) \\*
\quad \Longleftrightarrow \quad
\Big( \cev{\ct{\hat{c}}_{x_1} \cdots \ct{\hat{c}}_{x_n} \hat{c}_{y_1} \cdots \hat{c}_{y_n}} = 0 \quad \forall \ n \geq 2 \Big) .
}

Therefore, we are led to study the relaxation of static local connected $2n$-point correlation functions:
\eq{\label{eq:nn_conn_2n_func}
\cev{\ct{\hat{c}}_{x_1}\!(t) \ct{\hat{c}}_{x_2}\!(t) \cdots \ct{\hat{c}}_{x_n}\!(t) \hat{c}_{y_1}\!(t) \hat{c}_{y_2}\!(t) \cdots \hat{c}_{y_n}\!(t)} .
}
These functions, and their dynamic brethren, will be the primary objects of study in this paper.

We can now state precisely the \emph{cluster decomposition} condition that the initial state $\hat{\rho}_0$ is assumed to satisfy (in this section). We assume that
\eqal{\label{eq:nn_exp_cluster_decomp}
\cev{\ct{\hat{c}}_{x_1} \cdots \ct{\hat{c}}_{x_n} \hat{c}_{x_{n+1}} \cdots \hat{c}_{x_{2n}}} = \ &o(e^{-\abs{x_i - x_j}/\xi}) \nonumber \\*
&\text{as} \ \ \abs{x_i - x_j} \to \infty ,
}
for any pair of indices $i,j \in 1,2,\dots,2n$, where $\xi$ is some finite correlation length, and where ``$f(x) = o(g(x))$ as $x \to a$'' means that $f(x)/g(x) \to 0$ as $x \to a$. In the general treatment of Section~\ref{sec:setup} onwards, we will significantly weaken this assumption, and require only \emph{algebraic} decay (rather than exponential) of the initial connected correlation functions (Eq.~(\ref{eq:nn_exp_cluster_decomp}) will be replaced by Eq.~(\ref{eq:cluster_decomposition_def})).

\subsection{The single-particle propagator}\label{sec:nn_propagator}

Because $\hat{H}_0$ is quadratic, the fermion operators
\eq{
\hat{c}_x(t) = e^{i \hat{H}_0 t} \hat{c}_x e^{-i \hat{H}_0 t}
}
evolve linearly; they obey
\eq{\label{eq:nn_c(t)}
\hat{c}_x(t) = \sum_y G_{xy}(t) \hat{c}_y ,
}
where
\eq{
G_{xy}(t) = \frac{1}{L} \sum_k e^{ik(x-y) + it \cos{k}} .
}
It follows from Eq.~(\ref{eq:nn_c(t)}) that
\eq{
\mel{0}{\hat{c}_x(t) \hat{c}^{\dagger}_y(0)}{0} = G_{xy}(t) ,
}
where $\ket{0}$ is the fermion vacuum, specified by
\eq{
\hat{c}_x \ket{0} = 0 \quad \forall \ x .
}
Thus, $G_{xy}(t)$, which is defined as the coefficient appearing in Eq.~(\ref{eq:nn_c(t)}), may be identified as the single-particle propagator (the amplitude for a particle added to the vacuum at site $y$ to be found after time $t$ at site $x$).

Equation~(\ref{eq:nn_c(t)}) also implies that
\eq{
\text{Tr}\big(\comm{\hat{c}_x(t)}{\hat{c}^{\dagger}_y(0)}_+ \, \hat{\rho}_0 \big) = G_{xy}(t) ,
}
where $\comm{\hat{a}}{\hat{b}}_+ \coloneqq \hat{a} \hat{b} + \hat{b} \hat{a}$ denotes the anticommutator.
Thus, $G_{xy}(t)$ can also be identified with the retarded single-particle Green's function (if the hamiltonian is quadratic, this quantity is independent of the state $\hat{\rho}_0$).

\begin{figure}[t]
	\centering
	\includegraphics[width=\linewidth]{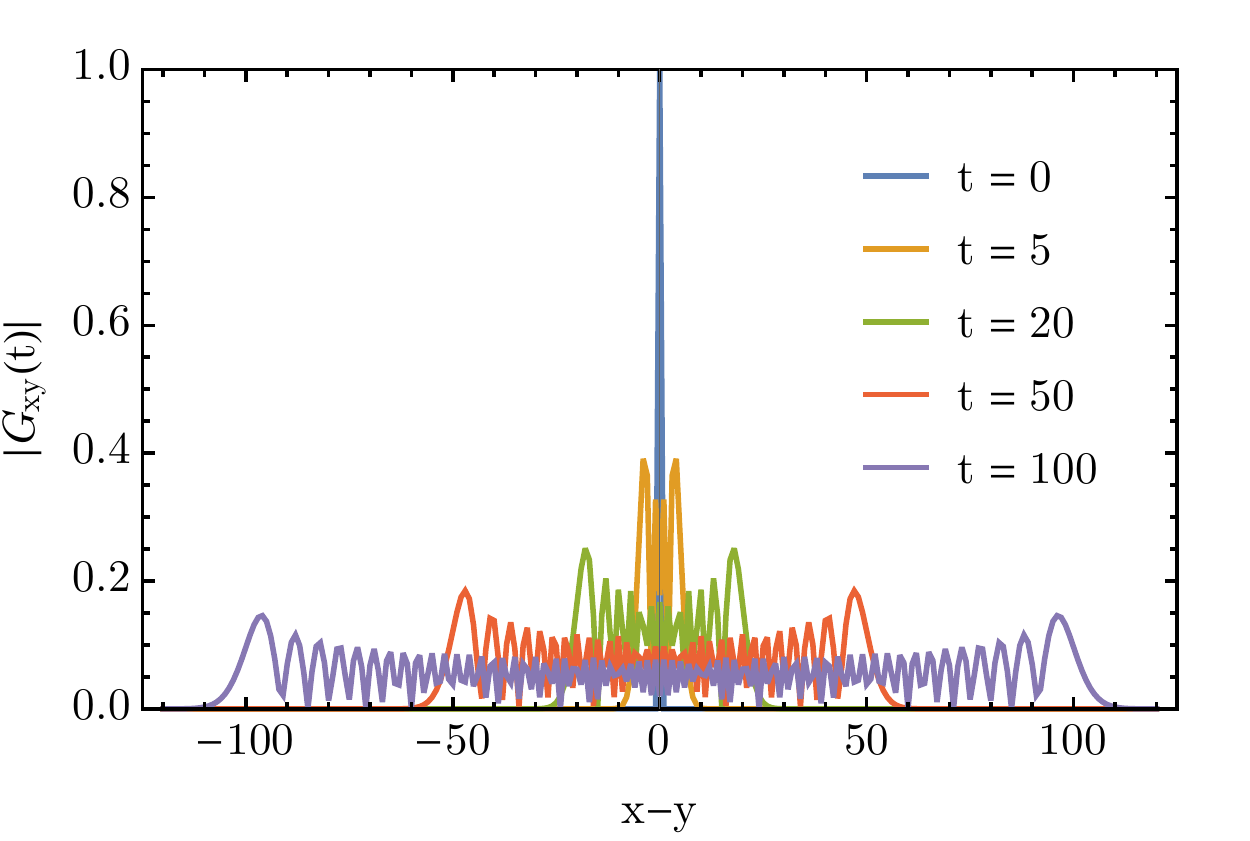}
	\caption{Magnitude of the single-particle propagator $\abs{G_{xy}(t)} = \abs{J_{x-y}(t)}$ for the model described by Eq.~(\ref{eq:nn_H0}).}
	\label{fig:bessel}
\end{figure}

These interpretations are useful for guessing properties of $G_{xy}(t)$ in situations in which one cannot write down a simple expression for it. We will not need to rely on intuition in this section, however. In the limit $L \to \infty$, one has
\eq{\label{eq:nn_Gt_int}
G_{xy}(t) = \int_{-\pi}^{\pi} \frac{\dd{k}}{2\pi} e^{i(x-y)k + it \cos{k}}
= i^{x-y} J_{x-y}(t) ,
}
where $J_n(z)$ is the Bessel function of order $n \in \mathbb{Z}$. The magnitude of the propagator, $\abs{G_{xy}(t)}$, is plotted for various values of $t$ in Figure~\ref{fig:bessel}.

We will show in the following that the leading late-time behavior of the connected $(2n \geq 4)$-point functions (\ref{eq:nn_conn_2n_func}) is actually determined by very basic properties of the propagator. Basically, all that matters is how the propagator ``spreads out'' with time. Let us characterize this ``spreading out'' more precisely.

Although $\abs{G_{xy}(t)}$ itself is a rapidly oscillating function of $x-y$ at fixed $t$, its smooth envelope is nonzero and slowly varying inside the ``lightcone'' $\abs{x-y} < t$, and decays exponentially to zero for $\abs{x-y} > t$. Qualitatively, this can be seen by glancing at Figure~\ref{fig:bessel}. More quantitatively, one can apply the method of stationary phase \cite{Stein1993} to the integral expression in Eq.~(\ref{eq:nn_Gt_int}) to obtain
\begin{widetext}
\eq{\label{eq:nn_Gt_asym}
\abs{G_{xy}(t)} \sim \left[ \frac{4}{\pi^2 (t^2 - r^2)} \right]^{1/4} 
\abs{\cos(\tfrac{\pi}{2}(r-\tfrac{1}{2}) - r\arcsin(r/t) - \sqrt{t^2-r^2})}
\quad \ \text{as} \quad t \to \infty 
\quad \text{if} \quad \frac{\abs{r}}{t} < 1 - O(t^{-1/3}) ,
}
and
\eq{
\abs{G_{xy}(t)} = o(t^{-n}) \quad \forall \ n
\quad \ \text{as} \quad t \to \infty 
\quad \text{if} \quad \frac{\abs{r}}{t} > 1 + O(t^{-1/3}) ,
}
where $r=x-y$. The $\abs{\cos(\cdots)}$ factor in Eq.~(\ref{eq:nn_Gt_asym}) describes the lattice-scale oscillations of $\abs{G_{xy}(t)}$; we replace it with a constant to obtain the smooth envelope. 
\end{widetext}

The two relevant properties of the propagator are that the interval of $\abs{x-y}$ values over which $G_{xy}(t)$ is non-negligible grows linearly with $t$, and that the matrix elements of $G_{xy}(t)$ inside this interval have a typical magnitude $\propto t^{-1/2}$. The second property can be extracted from Eq.~(\ref{eq:nn_Gt_asym}), but it can also be deduced very simply from the first property, as follows. Unitarity of time-evolution implies that $G(t)$ is a unitary matrix:
\eq{
1 = \sum_y \abs{G_{xy}(t)}^2 .
}
We can restrict the sum to the interval over which $\abs{G_{xy}(t)}$ is non-negligible:
\eq{
1 \approx \sum_{y \, \approx \, x-t}^{x+t} \abs{G_{xy}(t)}^2 .
}
Since the envelope of $\abs{G_{xy}(t)}$ is nonzero and slowly varying within this interval, and since the interval grows linearly with $t$, one must have $\abs{G_{xy}(t)} \sim t^{-1/2}$.

\subsection{Decay of local connected $(n \geq 4)$-point functions. ``Gaussification''}\label{sec:nn_gaussification}

\begin{figure}
	\centering
	\includegraphics[width=\linewidth]{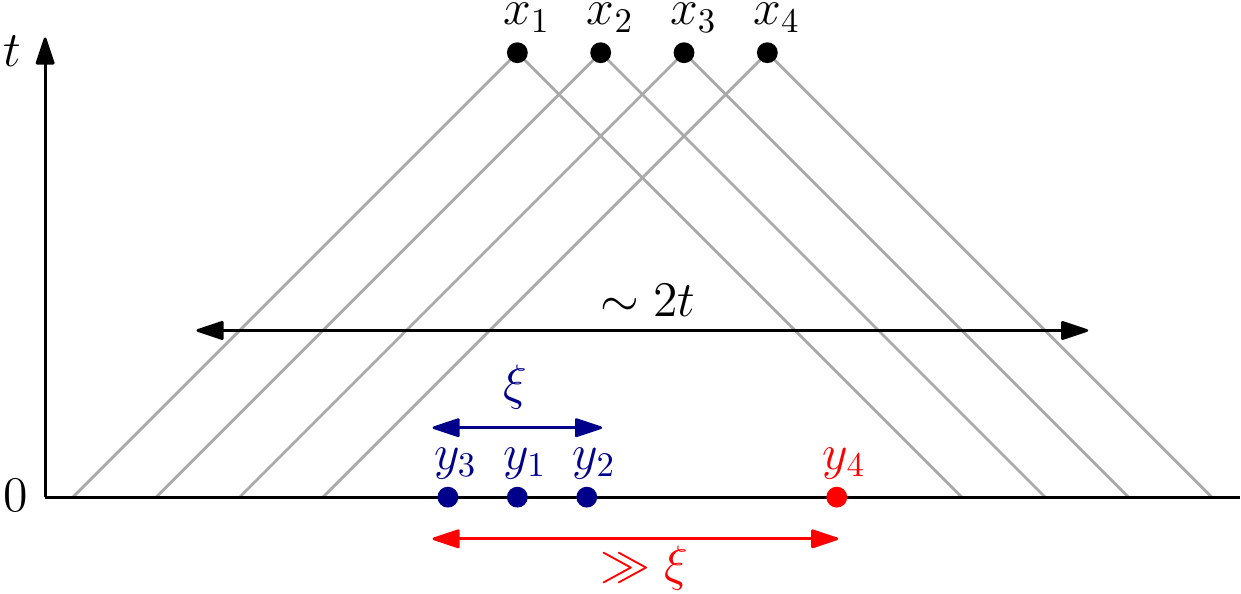}
	\caption{Schematic showing spreading of operators in the model of Eq.~(\ref{eq:nn_H0}), and how this leads to the decay of connected correlation functions as time $t$ increases. The points $x_j$ are the locations of the operators on the left side of Eq.~(\ref{eq:nn_4pt_connected}); the points $y_j$ are for a representative term in the sum on the right side of this equation. Each $y_j$ must lie inside the backward light-cone of $x_j$ in order for the propagator $G_{x_j y_j}\!(t)$ to be nonzero. Configurations of the $y$'s in which the distance between any pair is much greater than the correlation length $\xi$ (as is the case in the figure) give negligible contributions due to clustering of correlations in the initial state. This effectively restricts the sum over $y_1 \cdots y_4$ to a region of size $\sim \xi^3 t$ (Eq.~(\ref{eq:nn_4pt_sum_volume})).}
	\label{fig:spreading_1d}
\end{figure}

We are now in a position to understand why Wick factorization is recovered as $t$ increases. Consider the equal-time connected $4$-point function $\cev{\ct{\hat{c}}_{x_1}\!(t) \ct{\hat{c}}_{x_2}\!(t) \hat{c}_{x_3}\!(t) \hat{c}_{x_4}\!(t)}$. Equation~(\ref{eq:nn_4pt_conn_def}) shows that this function measures the extent to which the $4$-point function $\ev{\ct{\hat{c}}_{x_1}\!(t) \ct{\hat{c}}_{x_2}\!(t) \hat{c}_{x_3}\!(t) \hat{c}_{x_4}\!(t)}$ fails to Wick factorize. Using Eq.~(\ref{eq:nn_c(t)}) and its adjoint to express the operators at time $t$ in terms of operators at time zero,
\eqal{\label{eq:nn_4pt_connected}
\cev{\ct{\hat{c}}_{x_1}\!(t) \ct{\hat{c}}_{x_2}\!(t) \hat{c}_{x_3}\!(t) \hat{c}_{x_4}\!(t)} \vphantom{\sum} \nonumber \\*
= \sum_{y_1 \cdots y_4} \Big[
&G^*_{x_1 y_1}\!(t) G^*_{x_2 y_2}\!(t) G_{x_3 y_3}\!(t) G_{x_4 y_4}\!(t) \nonumber \\*[-0.5em]
&\times \cev{\ct{\hat{c}}_{y_1} \ct{\hat{c}}_{y_2} \hat{c}_{y_3} \hat{c}_{y_4}} \Big] .
}
We can estimate the magnitude of this quantity by multiplying the number of significant terms in the sum by the typical magnitude of each one. 
We have already seen that $\abs{G_{xy}(t)}$ is negligible outside the lightcone $\abs{x-y} \sim t$, and that it has typical magnitude $\abs{G_{xy}(t)} \sim t^{-1/2}$ inside. By our assumption (\ref{eq:nn_exp_cluster_decomp}) on exponential clustering of correlations in the initial state, the function $\cev{\ct{\hat{c}}_{y_1} \ct{\hat{c}}_{y_2} \hat{c}_{y_3} \hat{c}_{y_4}}$ is negligible whenever $\abs{y_i - y_j} > \xi$, where $\xi$ is the finite correlation length. As a result, the sum over $\mathbf{y} = (y_1 \cdots y_4)$ in Eq.~(\ref{eq:nn_4pt_connected}) is restricted to a region of size $\sim \xi^3 t$ (this is illustrated in Figure~\ref{fig:spreading_1d}):
\eqal{\label{eq:nn_4pt_sum_volume}
\text{Vol}\big\{ \mathbf{y} \big| \big[ &G^*_{x_1 y_1}\!(t) G^*_{x_2 y_2}\!(t) G_{x_3 y_3}\!(t) G_{x_4 y_4}\!(t) \nonumber \\*
&\times \cev{\ct{\hat{c}}_{y_1} \ct{\hat{c}}_{y_2} \hat{c}_{y_3} \hat{c}_{y_4}} \big] \ \text{non-negligible} \big\}
\sim \, \xi^3 t ,
}
while each term in the sum is of order
\eq{\label{eq:nn_4pt_summand}
\abs{G_{xy}(t)}^4 \sim t^{-2} .
}
Hence the right hand side of Eq.~(\ref{eq:nn_4pt_connected}) is of order $\sim t^{-1}$ and the connected function on the left vanishes in this manner as $t \to \infty$.

As $t$ increases, there will be additional constructive or destructive interference between different terms in the sum of Eq.~(\ref{eq:nn_4pt_connected}), that we have not taken into account in our crude accounting. Thus, we expect in general that
\eq{
\cev{\ct{\hat{c}}_{x_1}\!(t) \ct{\hat{c}}_{x_2}\!(t) \hat{c}_{x_3}\!(t) \hat{c}_{x_4}\!(t)} \sim \frac{z(t)}{t}
\quad \ \text{as} \quad t \to \infty ,
}
where $z(t)$ is some oscillatory function of time.

A similar argument shows that
\eq{\label{eq:nn_2n_pt_relaxation}
\cev{\ct{\hat{c}}_{x_1}\!(t) \cdots \ct{\hat{c}}_{x_n}\!(t) \hat{c}_{x_{n+1}}\!(t) \cdots \hat{c}_{x_{2n}}\!(t)}
\sim \frac{z_n(t)}{t^{n/2-1}} \vphantom{\bigg|}
}
as $t \to \infty$, where the $z_n(t)$ are some other oscillatory functions of time.

Thus, as $t \to \infty$, only the fully disconnected parts of local correlation functions survive (the parts that factorize into products of $2$-point functions); in other words, we recover Wick factorization as $t \to \infty$. As mentioned above, this is the defining property of a gaussian density matrix. We conclude that, as $t \to \infty$, the ``local state of the system'' may be described by a density matrix of the form
\eq{\label{eq:nn_gaussian_rho}
\hat{\rho}_1(t) = \frac{1}{Z_1(t)} \exp( - \sum_{x,y} \hat{c}^{\dagger}_x K_{xy}(t) \hat{c}_y ) ,
}
where $K_{xy}(t)$ is chosen such that
\eq{\label{eq:nn_Kt_constraint}
\Tr( \hat{c}^{\dagger}_x(t) \hat{c}_y(t) \hat{\rho}_1(t) ) = \ev{ \hat{c}^{\dagger}_x(t) \hat{c}_y(t) }
} 
for all sites $x,y$ with $\abs{x-y}$ finite in the limit $L \to \infty$, and $Z_1(t)$ ensures normalization. As long as $K_{xy}(t)$ is chosen to satisfy this condition at each time $t$ (actually, Eq.~(\ref{eq:nn_Kt_constraint}) only needs to hold up to terms of order $\sim t^{-1}$), we have
\eq{\label{eq:nn_Ot}
\ev{ \hat{\mathcal{O}}(t) } \sim \text{Tr} \big( \hat{\mathcal{O}}(t) \hat{\rho}_1(t) \big) + O(t^{-1})
\quad \ \text{as} \quad t \to \infty ,
} 
for all local observables $\hat{\mathcal{O}}$.

We have shown that the state becomes ``locally gaussian'' at late times. Following Ref.~\cite{Gluza2016}, we refer to this process as ``gaussification''. In Section~\ref{sec:gaussification} we describe gaussification in arbitrary quadratic lattice models by generalizing the chain of reasoning leading from Eq.~(\ref{eq:nn_c(t)}) to Eq.~(\ref{eq:nn_2n_pt_relaxation}).

\subsection{Equilibration of the local 2-point function to its GGE value}\label{sec:nn_2pt_relaxation}

It remains to compute the local equal-time 2-point function, and to verify that it relaxes to its stationary GGE value. By definition of the GGE, Eq.~(\ref{eq:nn_gge_rho}), this stationary value is
\eq{\label{eq:nn_2pt_gge}
\ev{\hat{c}^{\dagger}_{x_1} \hat{c}_{x_2}}_{\text{GGE}}
= \frac{1}{L} \sum_k e^{-ik(x_1 - x_2)} \ev{\hat{n}(k)} ,
}
where $\ev{\hat{n}(k)}$ is the expectation of the mode occupation number $\hat{n}(k)$ in the initial state.

The results of this subsection depend on a third assumption about the initial state $\hat{\rho}_0$, in addition to Eqs.~(\ref{eq:nn_N_cons}) and (\ref{eq:nn_exp_cluster_decomp}). Roughly speaking, we want to exclude situations in which the initial profiles of local conserved densities are inhomogeneous on length scales comparable to the system size---for instance, an initial state in which sites $x =1,2, \dots, L/2$ are occupied by fermions and the rest are empty. True local equilibration in such cases occurs on timescales of order $L$, simply because that is how long it takes a locally conserved density to flow across the system.

In order to formulate this assumption precisely, recall that the local conserved quantities (Eqs.~(\ref{eq:nn_I1m}) and (\ref{eq:nn_I2m})) are of the form
\eq{
\hat{I}_m = \sum_{x=1}^L \hat{\mathcal{I}}_{m,x} ,
}
where the density $\hat{\mathcal{I}}_{m,x}$ is supported on a finite interval of length $\lfloor{m/2}\rfloor$ near site $x$. Define the ``local excess density''
\eq{
\delta \mathcal{I}_m(x_0; L_0) \coloneqq \frac{1}{L_0} \sum_{x = x_0 - L_0/2}^{x_0 + L_0/2} \! \ev{\hat{\mathcal{I}}_{m,x}}
\, - \frac{1}{L} \ev{\hat{I}_m} .
}
We assume that these excess densities can be made small by taking $L_0$ sufficiently large (but finite and independent of $L$ as $L \to \infty$):
\eq{\label{eq:nn_homogeneous}
\exists \, L_0  : \, \delta \mathcal{I}_m(x_0; L_0) = O\Big(\frac{1}{L_0}\Big) \ \forall \ x_0, m 
\ \ \, \text{as} \ \ L \to \infty . 
\vphantom{\Big|}
}

We emphasize that the results of the previous subsections hold even when this assumption is violated. In particular, the system still ``gaussifies'' as described in Section~\ref{sec:nn_gaussification}. Thus, if the initial state violates Eq.~(\ref{eq:nn_homogeneous}), the natural description of the local state of the system at late times is in terms of a time-dependent gaussian density matrix, given by Eqs.~(\ref{eq:nn_gaussian_rho}) and (\ref{eq:nn_Kt_constraint}).

To study relaxation of the 2-point function, we will finally need to use the diagonal form (\ref{eq:nn_H0_diag}) of the hamiltonian $\hat{H}_0$, or equivalently, the full form (\ref{eq:nn_c(t)}) of the propagator $G_{xy}(t)$. We may write
\eqal{\label{eq:nn_2pt}
\ev{ \ct{\hat{c}}_{x_1}\!(t) \hat{c}_{x_2}\!(t) }
= \frac{1}{L} \sum_{k_1, k_2} \Big[
&e^{-i(k_1 x_1 - k_2 x_2)} e^{- i(\cos{k_1} - \cos{k_2}) t} \nonumber \\*[-0.5em]
&\times F(k_1, k_2) \Big] ,
}
where
\eq{\label{eq:nn_F}
F(k_1, k_2) \coloneqq \ev{ \ct{\hat{c}}\!(k_1) \hat{c}(k_2) } .
}

We begin by showing that, under the assumptions we have made, the function $F$ must have the form
\eqal{\label{eq:nn_F_form}
F(k_1, k_2) = \ &\delta_{k_1, k_2} \ev{\hat{n}(k_1)} \vphantom{\sum} \nonumber \\*
&+ \sum_{j=1}^{j_{\text{max}}} \delta_{k_1 - q_j, k_2} \, f_j(k_1)
+ \frac{1}{L} f(k_1, k_2) ,
}
where each $q_j \neq 0$ remains finite in the limit $L \to \infty$, and where $\ev{\hat{n}(k)}$, $f_j(k)$, and $f(k,k')$ are smooth $O(1)$ functions. The various Kronecker deltas contain all of the singular dependence of $F(k_1,k_2)$ on its arguments.

We arrive at Eq.~(\ref{eq:nn_F_form}) as follows. Invert the Fourier transformation and write
\eqal{\label{eq:nn_F_exp}
F(k_1, k_2) &= \frac{1}{L} \sum_{y_1, y_2} e^{i(k_1 y_1 - k_2 y_2)} \ev{ \ct{\hat{c}}_{y_1} \hat{c}_{y_2} } \nonumber \\*
&= \frac{1}{L} \sum_{y_1, y_2} \Big[
e^{i(k_1 + k_2) (y_1 - y_2)/2} \nonumber \\*[-0.5em]
&\qquad\qquad\quad \times e^{i(k_1 - k_2) (y_1 + y_2)/2} \ev{ \ct{\hat{c}}_{y_1} \hat{c}_{y_2} } \Big] .
}
The sums over $y_1$ and $y_2$ in Eq.~(\ref{eq:nn_F_exp}) may be performed with respect to the central coordinate $(y_1 + y_2)/2$ and relative coordinate $(y_1 - y_2)$. Due to clustering of correlations, the sum over the relative coordinate converges absolutely (it is effectively restricted to a finite window $\abs{y_1 - y_2} \lesssim \xi$), and consequently $F$ must be a smooth function of $(k_1 + k_2)$. On the other hand, the central coordinate is summed over the whole system, and so $F$ can depend in a singular manner on $(k_1 - k_2)$. In particular, $F(k_1,k_2)$ is $O(1)$ if and only if the terms in the sum over the central coordinate add constructively. This occurs when $(k_1 - k_2) = 0$ (in which case the phase factor in Eq.~(\ref{eq:nn_F}) is independent of the central coordinate), but it may also occur for $(k_1 - k_2) = q \neq 0$ if the initial state has a density wave with wavevector $q$, so that $\ev{ \ct{\hat{c}}_{y_1} \hat{c}_{y_2} } \propto e^{-iq(y_1 + y_2)/2}$. Our extra assumption on the initial state, Eq.~(\ref{eq:nn_homogeneous}), implies that $q \not\to 0$ as $L \to \infty$. This establishes the validity of Eq.~(\ref{eq:nn_F_form}).

Using Eq.~(\ref{eq:nn_F_form}) in Eq.~(\ref{eq:nn_2pt}), we obtain
\eqal{
\ev{ \ct{\hat{c}}_{x_1}\!(t) \hat{c}_{x_2}\!(t) } 
= \ &\frac{1}{L} \sum_{k} e^{-ik(x_1 - x_2)} \ev{ \hat{n}(k) } \nonumber \\*
&+ \sum_{j=1}^{j_{\text{max}}} \delta C^{(j)}_{x_1,x_2}\!(t) + \delta C_{x_1,x_2}\!(t) ,
}
where
\eqal{\label{eq:nn_dC(t)_j}
\delta C^{(j)}_{x_1,x_2}\!(t) = \frac{1}{L} \sum_{k} \Big[
&e^{-ik(x_1 - x_2) - iq_j x_2} \nonumber \\*[-0.5em]
&\times e^{- i[\cos{k} - \cos(k-q_j)] t} \, f_j(k) \Big] ,
}
and 
\eqal{\label{eq:nn_dC(t)}
\delta C_{x_1,x_2}\!(t) = \frac{1}{L^2} \sum_{k_1, k_2} \Big[
&e^{-i(k_1 x_1 - k_2 x_2)} e^{- i(\cos{k_1} - \cos{k_2}) t} \nonumber \\*[-0.5em]
&\times f(k_1, k_2) \Big] .
}

When $t \gg 1$, we may apply the method of stationary phase to estimate the time-dependent pieces. The cleanest way to do this is to first take $L \to \infty$, so that $\frac{1}{L} \sum_k \to \int \frac{\dd{k}}{2\pi}$, and only then take $t$ large, and that it what we will do here. However, we note in passing that it is also possible to perform a similar analysis \emph{without} first taking $L \to \infty$; one can use the Poission summation formula to represent $\frac{1}{L} \sum_k$ as a sum of integrals---each integral corresponding to a translated copy of the finite system---and then estimate each of these integrals by stationary phase. As long as $t < L/ v_{\text{max}}$, where $v_{\text{max}}$ is the maximal group velocity of particles in the system, the extra translated integrals generate only exponentially small (in $t$) corrections to the $L \to \infty$ result.

For completeness, let us briefly review the method of stationary phase. This method is described in detail in many standard texts, such as Ref.~\cite{Stein1993}. A nice heuristic and mathematically elementary treatment may be found in section 3.3 of Ref.~\cite{Mandel1995}.
 In the limit $L \to \infty$, Eqs.~(\ref{eq:nn_dC(t)_j}) and (\ref{eq:nn_dC(t)}) are both of the general form
\eq{
I(t) = \int \frac{\dd[d]{k}}{(2\pi)^d} \, a(k) \, e^{i \varphi(k) t} ,
}
where $a$ and $\varphi$ are smooth functions. The $k$-integral is one-dimensional in Eq.~(\ref{eq:nn_dC(t)_j}) and two-dimensional in Eq.~(\ref{eq:nn_dC(t)}). In both cases, the integral is over a compact region without boundary. As $t \to \infty$, the dominant contributions to the $I(t)$ integral come from the vicinity of points $k_*$ at which $\nabla_k \, \varphi(k_*) = 0$, called \emph{critical points} of $\varphi$. A critical point $k_*$ is \emph{nondegenerate} if the Hessian matrix at that point,
\eq{
\mathbf{H}_{ab}(k_*) \coloneqq \pdv{k_a} \pdv{k_b} \varphi(k_*) ,
}
is invertible. Each isolated nondegenerate critical point $k_j$ gives a contribution $I_j(t)$ to $I(t)$ that can be obtained (to leading order in $t$) by expanding the phase function $\varphi(k)$ up through quadratic order in $(k - k_j)$, extending the limits of the $k$-integral to infinity, and performing the resulting gaussian integral; the result is
\eq{
I_j(t) = \frac{e^{i(\pi/4)s_j}}{(2\pi t)^{d/2} \, \abs{\det \mathbf{H}(k_j)}^{1/2}} \, a(k_j) \, e^{i \varphi(k_j)t} + \cdots ,
}
where $s_j$ is the signature (number of positive eigenvalues minus number of negative eigenvalues) of the symmetric matrix $\mathbf{H}(k_j)$.
The dots are subleading terms proportional to higher derivatives of $a(k)$ evaluated at $k_j$.
Terms with $n$ derivatives are suppressed relative to the leading term by an additional factor of $t^{-n/2}$.

We obtain $I(t)$ by simply adding up these contributions (assuming $\varphi$ has no other critical points):
\eq{
I(t) \sim \sum_j I_j(t) .
}
Thus, whenever the phase function $\varphi$ has a finite number of critical points, all of which are nondegenerate (and assuming that the amplitude function $a(k)$ does not vanish at all of these points),
\eq{
I(t) \sim t^{-d/2} \quad \text{as} \quad t \to \infty .
}
This is the generic situation.

If, however, $\varphi$ does have \emph{degenerate} critical points, their contributions must also be accounted for.
The power of $t$ associated with such a contribution can often be estimated very simply as follows.
Assume that $k_*$ is a critical point at which $\varphi(k) - \varphi(k_*)$ has a zero of order $m$, while $a(k)$ has a zero of order $n$ (that is, the Taylor expansions of these functions about $k = k_*$ start with monomials of order $m$ and $n$ respectively). 
In spherical coordinates centered at $k_*$, we would have $\varphi(k_* + k) \approx \varphi(k_*) + \abs{k}^m \Phi(\theta)$ and $a(k_* + k) \approx \abs{k}^n A(\theta)$, where $\Phi$ and $A$ are appropriate functions of the angular variables, collectively denoted $\theta$.
Thus the leading contribution from the critical point is of the form
\eq{
I_j(t) \sim \int \frac{k^{d-1} \dd{k}}{(2\pi)^d} \int \dd{\Omega} \, k^n A(\theta) \, e^{i k^m \Phi(\theta) t} .
}
Scaling $t$ out of the integral by changing integration variables to $p = t^{1/m} k$, we obtain the estimate
\eq{
\label{eq:I(t)_general_asymptotic}
I_j(t) \sim t^{-(d+n)/m} \quad \text{as} \quad t \to \infty .
}
Note that larger $m$ leads to slower decay.
Thus, in the (non-generic) case that $\varphi$ has degenerate critical points, the ``most degenerate'' of these will typically dominate the $t \to \infty$ behavior of $I(t)$.
This concludes our brief mathematical interlude.

For a given $q_j \neq 0$, the phase function $\varphi(k;q_j) = \cos(k-q_j) - \cos(k)$ appearing in Eq.~(\ref{eq:nn_dC(t)_j}) has precisely two distinct nondegenerate critical points: $k = k_{\pm} = \tfrac{1}{2} (q_j \pm \pi)$. Thus, assuming that $f_j(k)$ does not vanish at these points, the method of stationary phase yields
\eq{\label{eq:nn_dCjt}
\delta C^{(j)}_{x_1,x_2}\!(t) \sim t^{-1/2} 
\quad \text{as} \quad t \to \infty .
}
A similar analysis applies to Eq.~(\ref{eq:nn_dC(t)}). In this case, the phase function $\varphi(k_1,k_2) = \cos{k_2} - \cos{k_1}$ has precisely four distinct nondegenerate critical points: $(k_1,k_2) = (0,0)$, $(0,\pi)$, $(\pi,0)$, and $(\pi,\pi)$. 
Thus, assuming that $f(k_1,k_2)$ does not vanish at these points,
\eq{\label{eq:nn_dCt}
\delta C_{x_1,x_2}\!(t) \sim t^{-1}
\quad \text{as} \quad t \to \infty .
}
Note that the locations of the critical points in $k$-space are determined by the dispersion relation of the hamiltonian $\hat{H}_0$, whereas the functions $f_j(k)$ and $f(k_1,k_2)$ are determined by the initial state.
Therefore, these functions will only vanish at the critical points for special, fine-tuned, choices of the initial state.
We conclude that, for \emph{generic} initial states, as $t \to \infty$,
\eq{\label{eq:nn_2pt_relaxation}
\ev{ \ct{\hat{c}}_{x_1}\!(t) \hat{c}_{x_2}\!(t) } 
\sim \ev{\hat{c}^{\dagger}_{x_1} \hat{c}_{x_2}}_{\text{GGE}} + R_{x_1 x_2}(t) ,
}
where the remainder $R_{xy}(t)$ is of order $t^{-1/2}$ if the initial state has a density wave, i.e.~if $\ev{ \ct{c}\!(k) \hat{c}(k - q) }$ is sharply peaked at one or more nonzero wavevectors $q$, and is of order $t^{-1}$ if not.

We have now explicitly shown that, for any initial state $\hat{\rho}_0$ that satisfies Eqs.~(\ref{eq:nn_N_cons}), (\ref{eq:nn_exp_cluster_decomp}) and (\ref{eq:nn_homogeneous}), all local observables of the system relax to their values in the GGE (\ref{eq:nn_gge_rho}) as $t \to \infty$ under time evolution generated by $\hat{H}_0$. Furthermore, we have obtained the exponents of the power laws governing the relaxation processes. We have shown that if the initial state has a density wave, then we generically expect the system to relax first to a (time-dependent) gaussian state like $\sim t^{-1}$, and then to relax to the GGE like $\sim t^{-1/2}$.
In Section~\ref{sec:gaussification} we describe relaxation of the local 2-point function---and hence relaxation of a gaussified state to the GGE---in arbitrary quadratic models, by generalizing the chain of reasoning leading from Eq.~(\ref{eq:nn_2pt}) to Eq.~(\ref{eq:nn_2pt_relaxation}).

Although we derived them for the specific model of Eq.~(\ref{eq:nn_H0}), the relaxation exponents $1/2$ and $1$ are actually generic for quenches to clean quadratic fermion models in one dimension.
Different exponents may be obtained if the final hamiltonian is fine-tuned (so that the dispersion relation has degenerate critical points) and/or if the initial state is fine-tuned (so that the functions $f_j(k)$ and $f(k_1,k_2)$ vanish at the critical points).
For instance, Ref.~\cite{Jhu2017} studied parameter quenches in a dimerized chain and in the Kitaev model of a 1d spinless $p$-wave superconductor, and obtained parameter-dependent relaxation exponents for the 2-point function.
In all cases, however, the exponents can be associated to degenerate critical points and/or to the vanishing of $f_j(k)$ or $f(k_1,k_2)$ at the critical points, and their values agree with the simple estimate (\ref{eq:I(t)_general_asymptotic}) (the authors of \cite{Jhu2017} perform a more sophisticated steepest descent analysis to also obtain the prefactors).
Moreover, one can easily verify that generic small perturbations of the pre-quench state and post-quench hamiltonian cause the exponents to return to the parameter-independent values $1/2$ and $1$.

Finally, we briefly comment on relaxation from initial states that violate Eq.~(\ref{eq:nn_homogeneous}). One might still expect the conclusions of this section to apply \emph{locally}, so that the system relaxes as described above toward a ``local GGE'' in which the Lagrange multipliers are slowly varying functions of position and time. This ``local GGE'' would in turn relax---over timescales comparable to the system size---to the global GGE of Eq.~(\ref{eq:nn_gge_rho}), in a manner consistent with a generalized theory of hydrodynamics \cite{Castro-Alvaredo2016}. This is certainly a tempting picture, but because one cannot associate a timescale to local power-law relaxation, it is not immediately clear that such a description---based on separation of timescales---is self-consistent. We will not explore these questions further in this paper.

\section{General treatment. Setup and basic definitions}\label{sec:setup}

\subsection{System}

We consider a lattice system of fermions or bosons in $d$ dimensions, with one orbital per lattice site and $N$ sites in total (the generalization to multiple orbitals per site is straightforward, and merely complicates the bookkeeping). Let $\hat{\psi}^-_x$ and $\hat{\psi}^+_x = (\hat{\psi}^-_x)^{\dagger}$ denote the annihilation and creation operators respectively for the site at position $x$.

Although we work on a lattice, we believe that many of our arguments also apply in the continuum limit, if the symbols in the equations are reinterpreted correctly; in particular $\hat{\psi}^{\pm}_x$ should be regarded as the operator that creates or destroys a wavepacket at position $x$. With this in mind, we will also make statements about relaxation in systems of massless particles, etc.

\subsection{Initial state}

At time $t = 0$, the system is prepared in some non-equilibrium initial state represented by the density matrix $\hat{\rho}_0$. For the majority of this paper, the only condition that we impose on $\hat{\rho}_0$ is that it have the \emph{cluster decomposition} property \cite{Weinberg1995}:
\eqal{\label{eq:cluster_decomposition_def}
\cev{\hat{\psi}^{a_1}_{x_1} \hat{\psi}^{a_2}_{x_2} \cdots \hat{\psi}^{a_n}_{x_n} } 
= \ &o(\abs{x_i - x_j}^{-(d+\epsilon)}) \nonumber \\*
&\text{as} \ \ \abs{x_i - x_j} \to \infty
}
for any pair of indices $i,j \in 1,2,\dots,n$, where $\epsilon > 0$ is some positive real number. Here $\cev{\cdots}$ denotes the \emph{connected correlation function} or \emph{cumulant} of the operators $\hat{\psi}^{a_1}_{x_1} \cdots \hat{\psi}^{a_n}_{x_n}$ in the state $\hat{\rho}_0$ (the definition of connected correlation function is reviewed in Appendix~\ref{app:cumulants}). Equation~(\ref{eq:cluster_decomposition_def}) says that the connected function vanishes at least as rapidly as $\abs{x_i - x_j}^{-(d+\epsilon)}$ when $\abs{x_i - x_j} \to \infty$, for some $\epsilon > 0$. 
The cluster decomposition property ensures that correlations in the state $\hat{\rho}_0$ factorize as groups of operators are taken far away from one another, and it is quite reasonable from a physical standpoint.
The cluster decomposition property (in fact a stronger exponential version of it) has been \emph{rigorously proven} for large classes of initial states. These include ground states of interacting local hamiltonians with a spectral gap \cite{Nachtergaele2006,Hastings2006}, as well as thermal states of arbitrary short-ranged fermionic lattice systems at sufficiently high temperature \cite{Kliesch2014}.
We emphasize again that the initial state $\hat{\rho}_0$ need not be related in any way to the hamiltonian of the system. For instance, it can be the ground state or thermal state of some completely different \emph{interacting} hamiltonian; the only requirement is that it satisfy Eq.~(\ref{eq:cluster_decomposition_def}).

In Sections~\ref{sec:2pt_relaxation}, \ref{sec:loc_equilibration_to_gge} and \ref{sec:floquet_gge_relaxation}, we will require the initial state to satisfy a second condition, in addition to cluster decomposition. This extra assumption is \emph{needed in these three sections and nowhere else}, so we state it when it first becomes relevant, in Section~\ref{sec:2pt_relaxation}. In the rest of the paper, only Eq.~(\ref{eq:cluster_decomposition_def}) is assumed.

\subsection{Hamiltonian}

For $t > 0$, the evolution of the system is governed by a quadratic, possibly time-dependent, hamiltonian of the form
\eq{\label{eq:quadratic_H}
\hat{H}(t) = \sum_{x,y} \negthinspace \Big[ \hat{\psi}^+_x h_{xy}(t) \hat{\psi}^-_y + \tfrac{1}{2} \big( \hat{\psi}^+_x \Delta_{xy}(t) \hat{\psi}^+_y + \text{h.c.} \big) \Big] ,
}
where $h_{xy}^* = h_{yx}$ and $\Delta_{xy} = \pm \Delta_{yx}$ for bosons/fermions respectively. This is the most general possible form of a quadratic hamiltonian. The term involving $h$ accounts for hopping and on-site potentials, while the term involving $\Delta$ allows for pairing. In the bosonic case, we assume that any linear terms have been eliminated by appropriately shifting the operators, and that the quasiparticle spectrum of $\hat{H}(t)$ is positive-definite.

It is often convenient to organize the annihilation and creation operators into a $2N$-component column vector $\hat{\Psi}$. If one orders the sites in some manner from $1$ to $N$, and temporarily denotes the operators acting on site number $j$ by $\hat{\psi}^{\pm}_j$, then
\eq{
\hat{\Psi} = (\hat{\psi}^-_1, \hat{\psi}^-_2, \cdots, \hat{\psi}^-_N, \hat{\psi}^+_1, \hat{\psi}^+_2, \cdots, \hat{\psi}^+_N)^T .
}
The hamiltonian can then be written in the form (column vector times matrix times row vector):
\eq{\label{eq:matrix_H}
\hat{H}(t) = \tfrac{1}{2} \hat{\Psi}^{\dagger} \mathcal{H}(t) \hat{\Psi} + \text{constant} ,
}
where
\eq{
\mathcal{H}(t) = \mat{ h(t) & \Delta(t) \\ \pm \Delta^*(t) & \pm h^*(t) } ,
}
and where the plus (minus) signs apply to bosons (fermions). $\mathcal{H}(t)$ is a $2N \times 2N$ hermitian matrix whose blocks are the matrices $h = h^{\dagger}$ and $\Delta = \pm \Delta^T$ with components $h_{xy}$ and $\Delta_{xy}$ (ordered to match the operators). In the bosonic case, we require $\mathcal{H}(t)$ to be positive-definite at each $t$ (this is equivalent to requiring the quasiparticle spectrum of $\hat{H}(t)$ to be positive-definite).

In general, we will refer to any $2N \times 2N$ matrix $\mathcal{M}$ as a \emph{canonical hermitian} matrix if it is of the form
\eq{\label{eq:canonical_hermitian_def}
\mathcal{M} = \mat{ X & Y \\ \pm Y^* & \pm X^* } , \qquad
\mathcal{M} = \mathcal{M}^{\dagger} .
}

We find it preferable to work in the Heisenberg picture throughout our analysis, so that the operators $\hat{\psi}^a_x(t)$ evolve with time $t$, while the unspecified density matrix $\hat{\rho}_0$ does not.

\subsection{Observables and relaxation}

The observables of interest are \emph{local correlation functions}; by this we mean any $n$-point function $\ev{\hat{\psi}^{a_1}_{x_1}(t_1) \cdots \hat{\psi}^{a_n}_{x_n}(t_n)}$ in which $\abs{x_i - x_j} \ll L$ for all pairs of indices $i,j \in 1,2,\dots,n$, where $L$ is the physical extent of the system (assumed to be of the same order of magnitude in each spatial direction). This notion of locality can be made precise in the thermodynamic limit $L \to \infty$, by requiring that all distances $\abs{x_i - x_j}$ remain finite.

We say that the system (whose true state in the Heisenberg picture is always given by $\hat{\rho}_0$) \emph{relaxes} to a state described by the density matrix $\hat{\rho}_1(t)$ if the latter reproduces all local correlation functions at late times.

\subsection{Gaussian density matrices}

A density matrix $\hat{\rho}$ is \emph{gaussian} if it is of the form
\eq{
\hat{\rho} = \frac{1}{Z} \exp( - \tfrac{1}{2} \hat{\Psi}^{\dagger} K \, \hat{\Psi} ) , 
}
where $K$ is a $2N \times 2N$ canonical hermitian matrix (that is, it satisfies Eq.~(\ref{eq:canonical_hermitian_def})). The quadratic form $\tfrac{1}{2} \hat{\Psi}^{\dagger} K \, \hat{\Psi}$ may be regarded as a ``statistical hamiltonian'' for the gaussian state (compare Eq.~(\ref{eq:matrix_H})). 

A density matrix $\hat{\rho}'$ is gaussian if and only if, for each $n \neq 2$, all connected $n$-point functions with respect to $\hat{\rho}'$ vanish (this is equivalent to Wick's theorem). Any gaussian state is therefore entirely determined by its $2$-point functions.

\section{``Gaussification'' of the initial state}\label{sec:gaussification}

We will first study the relaxation, in the sense defined above, of a system prepared in the initial state $\hat{\rho}_0$ and evolving according to the quadratic hamiltonian (\ref{eq:quadratic_H}), to a state described by a gaussian density matrix. Following Ref.~\cite{Gluza2016}, we refer to this process as ``gaussification''.

This section can be regarded as generalizing the logic that led from Eq.~(\ref{eq:nn_c(t)}) to Eq.~(\ref{eq:nn_gaussian_rho}) in Section~\ref{sec:nn_chain}.

\subsection{Spreading of operators. General properties of the propagator}\label{sec:propagator}

As stated earlier, we work in the Heisenberg picture. Since the hamiltonian (\ref{eq:quadratic_H}) is quadratic, the Heisenberg equations of motion for $\hat{\psi}^a_x(t)$ yield a system of \emph{linear} ordinary differential equations. These may be written in matrix form, following the notation of Eq.~(\ref{eq:matrix_H}), as
\eq{\label{eq:heisenberg_eom}
\pdv{t} \hat{\Psi}(t) = - i M(t) \hat{\Psi}(t) ,
}
where
\eq{
M(t) = \mat{h(t) & \Delta(t) \\ - \Delta^*(t) & -h^*(t)} .
}
Recall that $h^{\dagger} = h$ and $\Delta^T = \pm \Delta$ for bosons (fermions). Thus, for fermions, the matrix $M(t) = \mathcal{H}(t)$ is always hermitian, whereas for bosons it is only hermitian if $\Delta = 0$. In either case, one may immediately integrate this matrix differential equation to obtain
\eq{\label{eq:Psi(t)}
\hat{\Psi}(t) = G(t) \hat{\Psi}(0) ,
}
which defines the \emph{propagator} $G(t)$; in general $G(t)$ is the time-ordered exponential of the matrix-valued function $M(t)$:
\eq{
G(t) = \mat{G^{--}(t) & G^{-+}(t) \\ G^{+-}(t) & G^{++}(t) } = \mathcal{T} e^{-i \int_0^t M(t') \dd{t'}} .
}
One always has $G^{++}(t) = [G^{--}(t)]^*$ and $G^{-+}(t) = [G^{+-}(t)]^*$. The matrix $G(t)$ is unitary in the case of fermions (or bosons with $\Delta=0$), since in these cases it is the time-ordered exponential of a hermitian matrix-valued function. For bosons in general, $G(t)$ is instead \emph{pseudo-unitary}; it satisfies $G^{\dagger} \eta \, G = \eta$, where $\eta = I_N \oplus - I_N$ and $I_N$ is the $N \times N$ identity matrix. We will at first restrict attention to the cases in which $G(t)$ is unitary, and postpone the discussion of the slightly more subtle case of bosons with nonzero pairing (with $\Delta \neq 0$) to Section~\ref{sec:bosons}.

Equation~(\ref{eq:Psi(t)}) can be written in component form as
\eq{\label{eq:psi(t)}
\hat{\psi}^a_x(t) = \sum_{b=\pm} \sum_y G_{xy}^{ab}(t) \hat{\psi}^b_y ,
}
where $\hat{\psi}^b_y = \hat{\psi}^b_y(0)$. The components $G_{xy}^{ab}(t)$ of the propagator may be interpreted as giving the amplitude for a particle ($b = -$) or hole ($b = +$) added to the ``vacuum'' at position $y$ to be found, after time $t$ has elapsed, as a particle ($a = -$) or hole ($a = +$) at position $x$. $G_{xy}^{ab}(t)$ also equals the retarded single-particle Green's function of the system (both normal and anomalous parts); with a quadratic hamiltonian $\hat{H}(t)$, this Green's function is independent of the state $\hat{\rho}_0$.

Unitarity of the matrix $G(t)$ ensures that
\eq{\label{eq:G_sum}
\sum_{b=\pm} \sum_y \abs*{G_{xy}^{ab}(t)}^2 = 1 .
}
In accordance with the interpretation of $G_{xy}^{ab}(t)$ given above, this equation may be understood as expressing conservation of probability of particles along with holes.

Our argument for ``gaussification'' depends only on very coarse properties of the propagator---on whether and how rapidly it ``spreads'' as time progresses. Let us make these notions precise. Following the terminology used in Ref.~\cite{Gluza2016}, we say that the dynamics are \emph{delocalizing} at $(x,a)$ if
\eq{\label{eq:delocalizing_def}
\abs*{G_{xy}^{ab}(t)} \to 0
\quad \text{as} \quad t \to \infty
\quad \ \forall \ (y,b) ;
}
otherwise we say that the dynamics are \emph{localizing} at $(x,a)$.

If the dynamics are delocalizing at $(x,a)$, then for any $c > 0$, at sufficiently late times $t$ one has $\abs*{G_{xy}^{ab}(t)} < c$ for all $(y,b)$. In order to satisfy Eq.~(\ref{eq:G_sum}), $\abs*{G_{xy}^{ab}(t)}$ must then be nonzero for at least $1/c^2$ pairs $(y,b)$. Thus, ``delocalizing dynamics'' requires spreading of the propagator. In order to quantify how rapidly this spreading occurs, consider the smooth envelope $\wtl{G}_{xy}^{ab}(t)$ of $\abs*{G_{xy}^{ab}(t)}$, obtained by coarse-graining the latter in $x$ and $y$ (in the example of Section~\ref{sec:nn_chain}, for instance, we obtain $\wtl{G}_{xy}(t)$ by averaging the curves in Figure~\ref{fig:bessel} over their rapid oscillations on the lattice scale). For given position $x$, index $a$, time $t$, and constant $\delta > 0$, define
\eq{
\mathcal{D}_x^a(t; \delta) \coloneqq \big\{ y \, \big| \, \wtl{G}_{xy}^{a+}(t) > \delta \ \ \text{or} \ \ \wtl{G}_{xy}^{a-}(t) > \delta \big\} 
}
and
\eq{
\mathcal{V}_x^a(t; \delta) \coloneqq \text{Vol}_y[\mathcal{D}_x^a(t; \delta)] .
}
By choosing $\delta$ small enough, we can ensure that, to any desired accuracy,
\eq{\label{eq:G_sum_approx}
\sum_{y \in \mathcal{D}_x^a(t;\delta)} \, \sum_{b=\pm} \abs*{G_{xy}^{ab}(t)}^2 \eqqcolon 1 - \epsilon^2(t,\delta) \approx 1,
}
where the first equality defines $\epsilon(t,\delta)$. Thus, whenever $G_{xy}^{ab}(t)$ is present in a sum over $y$, we may restrict the sum to $y \in \mathcal{D}_x^a(t;\delta)$ while only making an error of order $\epsilon(t,\delta)$. In what follows, we will assume that $\delta = \delta_*(t)$ has been chosen small enough so that the error $\epsilon(t,\delta_*(t))$ is negligible, and suppress it in writing
\eq{
\mathcal{D}_x^a(t) = \mathcal{D}_x^a(t; \delta_*(t))
}
and
\eq{
\mathcal{V}_x^a(t) = \mathcal{V}_x^a(t; \delta_*(t)) .
}

In many cases of interest, including lattice systems with Lieb-Robinson bounds \cite{Lieb1972,Hastings2004,Gluza2016}, $\mathcal{V}_x^a(t; \delta)$ depends much more weakly on $\delta$ than does $\epsilon(t,\delta)$ in the limit $\delta \to 0$ (a glance back at Figure~\ref{fig:bessel} shows that this is true in the example of Section~\ref{sec:nn_chain}). In order to satisfy Eq.~(\ref{eq:G_sum_approx}), the non-negligible components $G^{ab}_{xy}(t)$, which belong to the region $y \in \mathcal{D}_x^a(t)$, must then have magnitude
\eq{
\abs*{G^{ab}_{xy}(t)} \sim [\mathcal{V}_x^a(t)]^{-1/2}
\quad \text{for typical} \ \, y \in \mathcal{D}_x^a(t) .
}
If the dynamics are delocalizing, one must have $\mathcal{V}_x^a(t) \to \infty$ as $t \to \infty$.

Usually, the dimension $d'$ of the region $\mathcal{D}_x^a(t)$ equals the dimension $d$ of the ambient space. However, there are also cases in which $d' < d$. For instance, for a system of massless particles with an isotropic dispersion relation, $\mathcal{D}_x^a(t)$ is the $d' = (d-1)$-dimensional surface of a $d$-dimensional sphere centered at $x$. 

Before proceeding, let us comment on three generic ways in which the dynamics may fail to delocalize:

\begin{enumerate}[leftmargin=*]

\item The most obvious one is that $\hat{H}$ describes a system that is Anderson localized \cite{Anderson1958}; in this case the dynamics are localizing at all points $x$. 

\item More generally, imagine that the quasiparticle spectrum of $\hat{H}$ includes a level whose wavefunction is exponentially localized in space near position $x_0$. The propagator $G^{ab}_{xy}(t)$ will then include a term, due to the localized state, that does not vanish as $t \to \infty$. However, this contribution will be exponentially small in the distances $\abs{x - x_0}$ or $\abs{y - x_0}$ if either of these is large. Thus, to an excellent approximaiton, the dynamics will only be localizing very near $x_0$, and will remain delocalizing elsewhere. We will discuss the special effects that arise when the quasiparticle spectrum of $\hat{H}$ contains one or more localized states, in addition to extended states, in Section~\ref{sec:localized_states}. For the remainder of the paper, we exclude this possibility. Because we define relaxation as a \emph{local} phenomenon, however, our general conclusions also apply to systems \emph{with} localized states, as long as we consider a region of space that is sufficiently far from them. 

\item Finally, consider a system of non-interacting particles moving in two dimensions in a constant perpendicular magnetic field. In this case, the dynamics are again localizing; the propagator $G(t)$ is a periodic function of time \cite{Feynman2010}. This may be inferred from the fact that, in the classical problem, all particles move in circular orbits at the cyclotron frequency $\omega_0 = eB/mc$, regardless of their initial velocity (here $m$ is the mass and $e$ the charge of each particle, $B$ is the magnitude of the magnetic field, and $c$ is the speed of light). Consequently, the wavefunction of a single particle prepared in a wavepacket at some point $\mathbf{r}_0$ simply expands and contracts rhythmically with period $2\pi/\omega_0$.

\end{enumerate}

\subsection{Decay of connected correlation functions. ``Gaussification''}\label{sec:n_function_decay}

Using Eq.~(\ref{eq:Psi(t)}), any time-dependent connected $n$-point function can be expressed as a linear combination of connected $n$-point functions at time zero:
\eqal{\label{eq:connected_n_function}
\cev{\hat{\psi}^{a_1}_{x_1}(t_1) \hat{\psi}^{a_2}_{x_2}(t_2) \cdots \hat{\psi}^{a_n}_{x_n}(t_n)} \vphantom{\sum} \nonumber \\*
=  \sum_{b_1 \cdots b_n = \pm} \ \sum_{y_1 \cdots y_n} \Big[
&G^{a_1 b_1}_{x_1 y_1}(t_1) \cdots G^{a_n b_n}_{x_n y_n}(t_n) \nonumber \\*[-0.5em]
&\times \cev{\hat{\psi}^{b_1}_{y_1} \hat{\psi}^{b_2}_{y_2} \cdots \hat{\psi}^{b_n}_{y_n}} \Big] . \quad
}
We are interested in \emph{local} correlation functions, so we assume that the $x_j$'s are all close to one another (relative to the size of the system). We can estimate the magnitude of the connected $n$-point function by simply multiplying the number of significant terms in the sum by the typical magnitude of each one. Based on the discussion in Section~\ref{sec:propagator}, the summand is negligible unless each $y_j$ is contained in the appropriate region $\mathcal{D}_{x_j}^{a_j}(t_j)$. Assume for a moment that the initial state $\hat{\rho}_0$ obeys a strong version of cluster decomposition, and has a finite correlation length $\xi$ such that $\cev{\hat{\psi}^{b_1}_{y_1} \hat{\psi}^{b_2}_{y_2} \cdots \hat{\psi}^{b_n}_{y_n}}$ is negligible whenever $\abs{y_i - y_j} \gg \xi$. Then, the summand at $\mathbf{y} = (y_1, y_2, \dots, y_n)$ is significant only if $\mathbf{y} \in \mathcal{D}(\{x_i,t_i\})$, where
\eqal{
\mathcal{D}(\{x_i,t_i\}) \approx \big\{ \mathbf{y} \, \big| \ &y_j \in \mathcal{D}_{x_j}^{a_j}(t_j) \ \, \forall \ j \nonumber \\*
&\text{and} \ \, \abs{y_i - y_j} < \xi \ \ \forall \ i,j \big\} ,
}
and the number of significant terms in the sum, $\mathcal{N}(t)$, is proportional to the volume, in $\mathbf{y}$-space, of $\mathcal{D}(\{x_i,t_i\})$.

\begin{figure*}
	\centering
	\includegraphics[width=0.35\textwidth]{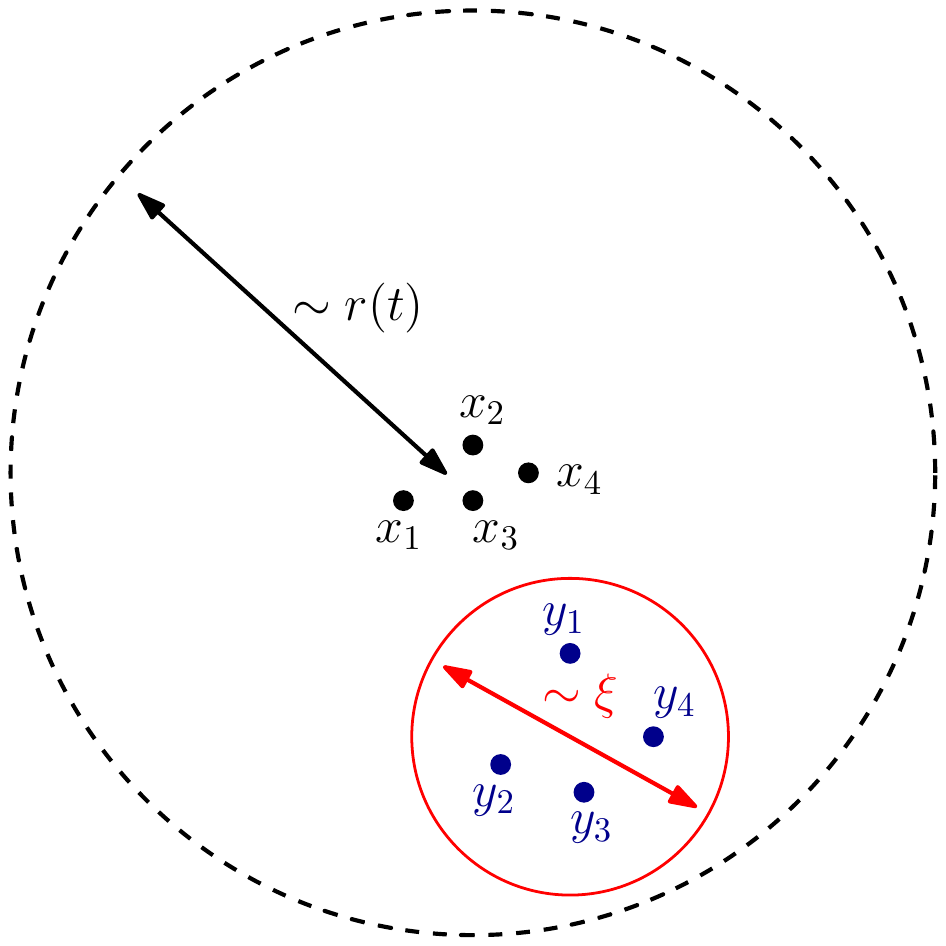} 
	\hspace{1.5em} \vrule \hspace{1.8em}
	\includegraphics[width=0.35\textwidth]{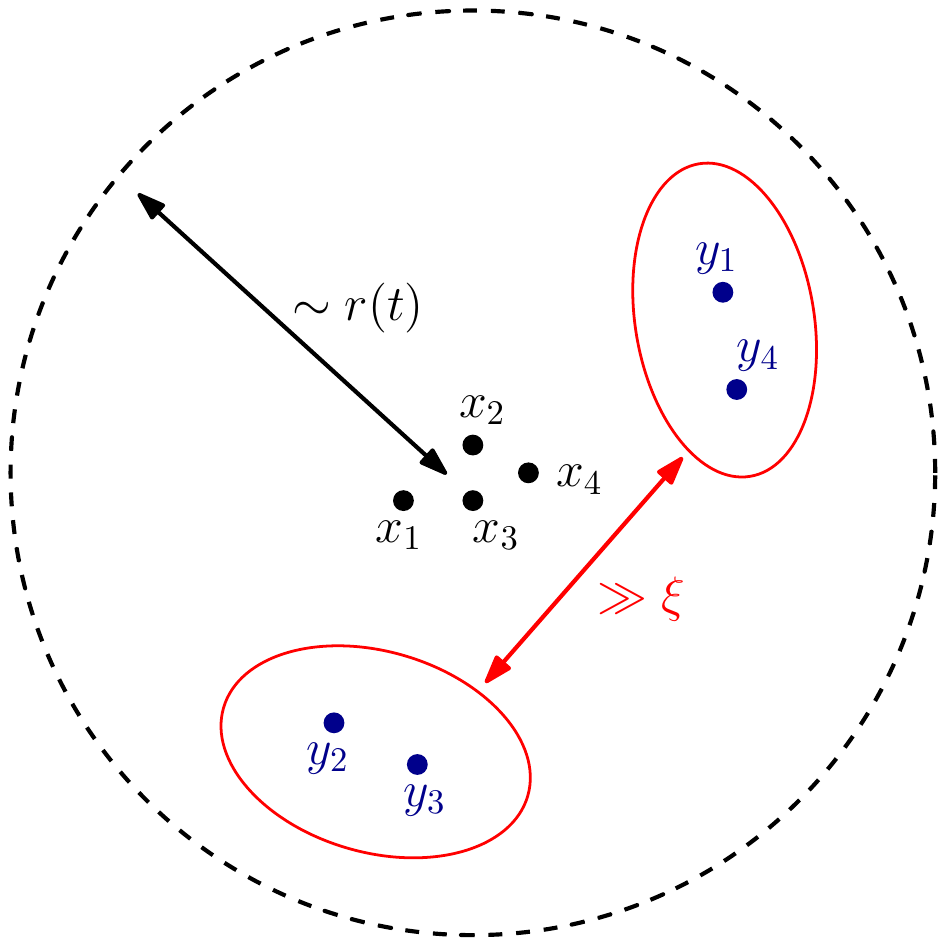}
	\caption{Schematic showing how the spreading of operators in $d=2$ dimensions causes the connected $4$-point function $\cev{\hat{\psi}^{a_1}_{x_1}(t) \hat{\psi}^{a_2}_{x_2}(t) \hat{\psi}^{a_3}_{x_3}(t) \hat{\psi}^{a_4}_{x_4}(t)}$ to decay as time $t$ increases. As in Eq.~(\ref{eq:connected_n_function}), this function is expressed as a weighted sum of connected $4$-point functions at time zero, $\cev{\hat{\psi}^{b_1}_{y_1} \hat{\psi}^{b_2}_{y_2} \hat{\psi}^{b_3}_{y_3} \hat{\psi}^{b_4}_{y_4}}$. Cluster decomposition ensures that only configurations of the $y$'s of the form depicted in the left panel contribute to the sum; configurations like that shown in the right panel \emph{do not}, because the connected function $\cev{\hat{\psi}^{b_1}_{y_1} \hat{\psi}^{b_2}_{y_2} \hat{\psi}^{b_3}_{y_3} \hat{\psi}^{b_4}_{y_4}}$ is negligible. This restriction in allowed phase space is ultimately responsible for the power-law decay of all connected $3$- and higher-point functions, as explained in the text.}
	\label{fig:2d_4pt_spreading}
\end{figure*}

With delocalizing dynamics, each region $\mathcal{D}_{x_j}^{a_j}(t_j)$ grows without bound as $t \to \infty$, so that $\mathcal{V}_{x_j}^{a_j}(t_j) \gg \xi^d$ at late times. In this case, it is easy to see that the number of significant terms in the sum is
\eq{
\mathcal{N}(t) \sim \mathcal{V}(t) \xi^{(n-1)d'} ,
}
where
\eq{\label{eq:V(t)_def}
\mathcal{V}(t) = \text{min}\{\mathcal{V}_{x_1}^{a_1}(t_1), \mathcal{V}_{x_2}^{a_2}(t_2), \cdots, \mathcal{V}_{x_n}^{a_n}(t_n) \} .
}
The factor of $\mathcal{V}(t)$ comes from a sum over the central coordinate $\bar{y} = \frac{1}{n} (y_1 + y_2 + \cdots + y_n)$, while the $(n-1)$ factors of $\xi^{d'}$ come from sums over the relative $y$-coordinates; the latter are restricted by cluster decomposition, while the former is not. This straightforward geometric argument is illustrated in Figure~\ref{fig:2d_4pt_spreading}.

Meanwhile, each $G^{ab}_{xy}(t_j)$ factor in the summand has typical magnitude $\sim [\mathcal{V}_x^a(t_j)]^{-1/2} \lesssim [\mathcal{V}(t)]^{-1/2}$. We conclude that $\abs*{\cev{\hat{\psi}^{a_1}_{x_1}(t_1) \hat{\psi}^{a_2}_{x_2}(t_2) \cdots \hat{\psi}^{a_n}_{x_n}(t_n)}} \lesssim [\mathcal{V}(t)]^{-(n/2 - 1)}$ as $t \to \infty$. If the various times $t_j$ are comparable (quantitatively, if the time differences $\abs{t_i - t_j}$ are small compared with the average time $\bar{t} = (t_1 + t_2 + \cdots + t_n)/n$), then we expect that $\mathcal{V}_{x_j}^{a_j}(t_j) \approx \mathcal{V}(t)$, and we may boldly promote this bound to an asymptotic estimate of the relaxation rate: $\cev{\hat{\psi}^{a_1}_{x_1}(t_1) \hat{\psi}^{a_2}_{x_2}(t_2) \cdots \hat{\psi}^{a_n}_{x_n}(t_n)} \sim [\mathcal{V}(t)]^{-(n/2 - 1)}$ as $t \to \infty$. This estimate ignores all interference between terms in the sum in Eq.~(\ref{eq:connected_n_function}). We briefly comment on some of these neglected interference effects at the end of this section.

With \emph{localizing} dynamics (as in quenches to disordered hamiltonians in $d=1$ or $2$ dimensions), the result depends crucially on the ratio of the localization length $\xi_{\text{loc}}$ to $\xi$. If $\xi_{\text{loc}} \gg \xi$, the conclusions of the previous paragraph are essentially unchanged, except that $\mathcal{V}(t) \to (\xi_{\text{loc}})^{d'}$ as $t \to \infty$. Thus, the connected functions still relax like $[\mathcal{V}(t)]^{-(n/2 - 1)}$, but to a finite value of order $\sim (\xi_{\text{loc}})^{-(n/2-1)d'}$, rather than to zero, and subsequently oscillate forever. If $\xi_{\text{loc}} < \xi$, then the $y$-sums in Eq.~(\ref{eq:connected_n_function}) are always restricted to regions of size $\sim (\xi_{\text{loc}})^{d'}$, cluster decomposition plays no role, and one expects little or no relaxation to occur. One can also consider the intermediate case in which the dynamics has both a localizing \emph{and} a delocalizing component. We study this in some detail in Section~\ref{sec:localized_states}.

A slight refinement of the argument just presented allows us to handle initial states in which the correlation length $\xi$ is infinite, but which nevertheless obey the weaker algebraic form of cluster decomposition (\ref{eq:cluster_decomposition_def}). Thus, assume that $\cev{\hat{\psi}^{b_1}_{y_1} \hat{\psi}^{b_2}_{y_2} \cdots \hat{\psi}^{b_n}_{y_n}} \sim \abs{y_i - y_j}^{-(d+\epsilon)}$ as $\abs{y_i - y_j} \to \infty$, with $\epsilon > 0$. Let $\xi$ now denote the length scale beyond which this power law is valid. Each propagator $G^{ab}_{xy}(t)$ factor in Eq.~(\ref{eq:connected_n_function}) still has typical magnitude $\lesssim [\mathcal{V}(t)]^{-1/2}$. We may rewrite the sum over $y_1 \cdots y_n$ as a sum over one central coordinate $\bar{y}$ and $(n-1)$ relative coordinates $z_j$. The sum over $\bar{y}$ is unrestricted by cluster decomposition, and yields a factor $\sim \mathcal{V}(t)$ as before. In order to estimate the sums over the relative coordinates, assume that each region $\mathcal{D}_{x_j}^{a_j}(t_j)$ is $d$-dimensional, and let $r(t)$ denote some typical length scale of these regions. Then,
\eqal{
\sum_{z_1 \cdots z_{n-1}} &\big\lvert \cev{\hat{\psi}^{b_1}_{y + z_1} \hat{\psi}^{b_2}_{y + z_2} \cdots \hat{\psi}^{b_n}_{y - (z_1 + z_2 + \cdots + z_{n-1})}} \big\rvert \nonumber \\*[-0.5em]
&\qquad\qquad \sim \xi^{(n-1)d} + \left( \int_{\xi}^{r(t)} \frac{\abs{z}^{d-1} \dd{\abs{z}}}{\abs{z}^{d+\epsilon}} \right)^{n-1} \nonumber \\*[0.4em]
&\qquad\qquad \sim \xi^{(n-1)d} + \xi^{-(n-1)\epsilon} / \epsilon,
}
where we have retained only the leading terms in the limit $r(t) \gg \xi$ (if the dynamics are delocalizing, $r(t) \to \infty$ as $t \to \infty$, so this limit will be reached at late times). The important point is that this leading term is a constant independent of $t$. Consequently, our earlier asymptotic estimate of the relaxation rate of the connected $n$-point function is not modified. If instead the regions $\mathcal{D}_{x_j}^{a_j}(t_j)$ are $d'$-dimensional (with $d' < d$), the requirement that each $y_j$ lie on the appropriate $d'$-dimensional manifold places some additional constraints on the $z_j$'s, but this is a detail that does not affect the main conclusion. 

Thus, whenever the initial state obeys cluster decomposition, as defined in Eq.~(\ref{eq:cluster_decomposition_def}), we expect that
\eq{\label{eq:n_function_decay}
\big\lvert \cev{\hat{\psi}^{a_1}_{x_1}(t_1) \hat{\psi}^{a_2}_{x_2}(t_2) \cdots \hat{\psi}^{a_n}_{x_n}(t_n)} \big\rvert  
\sim [\mathcal{V}(t)]^{-(n/2-1)} ,
}
with $\mathcal{V}(t)$ given by Eq.~(\ref{eq:V(t)_def}). Our arguments suggest that this result holds whenever $\mathcal{V}(t) \gg \xi^{d'}$, where $d' \leq d$ is the effective dimension of the regions $\mathcal{D}_{x_j}^{a_j}(t_j)$, and $\xi$ is an appropriate length scale in the initial state (either the correlation length, if this is finite, or the length scale beyond which the initial connected $n$-point functions exhibit the power law decay required by cluster decomposition).

Notice that Eq.~(\ref{eq:n_function_decay}) does not give any information about the relaxation behavior of the $2$-point function, since the exponent of $\mathcal{V}(t)$ vanishes when $n = 2$. This is easily understood. As we saw in the example of Section~\ref{sec:nn_chain}, and as we will show later in generality, the relaxation of the $2$-point function is governed by interference between the terms in the sum in Eq.~(\ref{eq:connected_n_function}). This interference was completely ignored in our derivation of Eq.~(\ref{eq:n_function_decay}), which relied only on gross phase space arguments. For $n > 2$, we hypothesize that the neglected interference effects merely lead to an additional oscillatory time-dependence about the power-law decay exhibited in Eq.~(\ref{eq:n_function_decay}), \emph{without} modifying the exponent of the power law itself.

\subsection{Relaxation power laws}\label{sec:rates}

Equation~(\ref{eq:n_function_decay}) gives estimates of the leading time-dependence of all local $(n > 2)$-point connected functions of the system in terms of the coarse spreading behavior of the propagator (as encoded in the function $\mathcal{V}(t)$). Although a detailed study of possible spreading behaviors is beyond the scope of this paper, we describe some generic types of spreading below. 

Typical spreading behaviors fall into two broad classes. In the first class, which we call ``volume spreading'', the smooth envelope $\wtl{G}_{xy}^{ab}(t)$ of the propagator (as defined in Section~\ref{sec:propagator}) is non-negligible for most points $y$ inside a $d$-dimensional region of characteristic size $r(t)$ centered at position $x$, so that $\mathcal{V}(t) \sim [r(t)]^d$. In the second class, which we call ``area spreading'', $\wtl{G}_{xy}^{ab}(t)$ is non-negligible only for points $y$ near the $(d-1)$-dimensional surface of such a region of size $r(t)$ centered at $x$, so that $\mathcal{V}(t) \sim [r(t)]^{(d-1)}$. In either case, the dynamics are delocalizing if $r(t) \to \infty$ as $t \to \infty$, and localizing if not (with the exception of area spreading in $d=1$, a case that we discuss separately in Section~\ref{sec:d=1}). 

\begin{figure*}
	\centering
	\includegraphics[width=0.75\textwidth]{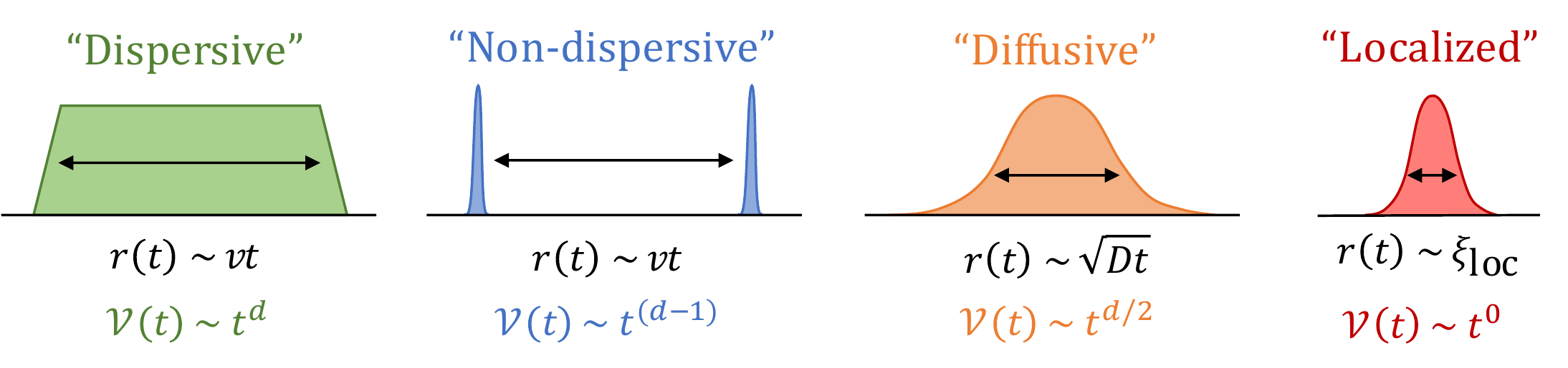} 
	\caption{Paradigmatic spreading behaviors of 1-particle propagators. This list is certainly \emph{not} exhaustive, but the spreading behaviors shown may be regarded as ``typical''.}
	\label{fig:propagators}
\end{figure*}

We expect behavior of the ``volume'' type for massive particles in a slowly varying potential (\emph{dispersive} spreading) and of the ``area'' type for massless particles (\emph{non-dispersive} spreading). In both these cases, $r(t) \sim v t$, where $v$ is the maximum local group velocity of the particles. In the presence of weak disorder, we again expect behavior of the ``volume'' type. In $d \geq 3$ dimensions, the expectation is \emph{diffusive} spreading of the form $r(t) \sim \sqrt{Dt}$ (here $D$ is the diffusion constant), while in $d = 1$ and $2$ dimensions the expectation is that $r(t)$ saturates at a finite localization length, $r(t) \sim \xi_{\text{loc}}$ as $t \to \infty$ \cite{Lee1985}. These four paradigmatic spreading behaviors are depicted schematically in Figure~\ref{fig:propagators}. The corresponding relaxation exponents may be easily obtained using Eq.~(\ref{eq:n_function_decay}).

For a time-independent hamiltonian $\hat{H}$ with a Lieb-Robinson bound \cite{Lieb1972,Hastings2004}, we expect the propagator to \emph{generically} behave in one of these manners; in the example of Section~\ref{sec:nn_chain}, for instance, the propagator exhibited what we are now calling dispersive spreading (of course, there are exceptions, such as the pathological ones noted in Section~\ref{sec:propagator}). More complicated behavior is certainly possible for time-dependent hamiltonians $\hat{H}(t)$, but generically we expect that these will still lead to spreading of either the ``volume'' or ``area'' types, with some characteristic size $r(t)$ that must be computed on a case-by-case basis. 

If the hamiltonian $\hat{H}$ contains non-local terms, so that there is no Lieb-Robinson bound, we cannot say as much about the envelope of the propagator. However, the unitarity condition (\ref{eq:G_sum}) still relates the typical magnitude of non-negligible matrix elements $G^{ab}_{xy}(t)$ to the volume $\mathcal{V}(t)$ of the region on which the propagator is meaningfully supported: $G^{ab}_{xy}(t) \sim [\mathcal{V}(t)]^{-1/2}$. Since, in the absence of a Lieb-Robinson bound, we expect $\mathcal{V}(t)$ to grow quite rapidly, the basic argument of Section~\ref{sec:n_function_decay} still applies, and we expect the system to ``gaussify'' rapidly (as measured by local operators).

Therefore, if the hamiltonian describes a delocalized system in the sense that $r(t) \to \infty$ as $t \to \infty$, then all local ($n \geq 3$)-point connected functions decay with the power laws obtained above, and the system can be described at late times by a gaussian density matrix. As mentioned earlier, there is one important exception to this result, which we now discuss.

\subsection{Non-dispersive spreading in $d=1$ dimension: absence of gaussification}\label{sec:d=1}

Recently, Sotiriadis \cite{Sotiriadis2016} has analytically studied the quench dynamics of a massless free bosonic scalar field in one spatial dimension, and has shown that the system always retains significant memory of non-gaussian initial correlations. Thus, the system fails to relax to the corresponding bosonic GGE, which is gaussian.
A very similar result was obtained earlier by Ngo Dinh et al.~\cite{NgoDinh2013}.

This result can be understood very easily within the framework that we have established above. The propagator of massless particles is supported entirely along the light cone. In $d = 1$ dimension, at each instant of time, the light cone simply consists of two points. Therefore, unitarity implies that the propagator can never decay; it follows from the analogue of Eq.~(\ref{eq:connected_n_function}) that higher connected correlation functions never relax to zero.

More generally, for any system whose propagator exhibits ``area spreading'', we have $\mathcal{V}(t) \sim [r(t)]^{(d-1)}$, and so $\abs{G_{xy}(t)} \sim [r(t)]^{-(d-1)/2}$. In $d=1$ dimension, these factors are constant, implying that higher connected correlation functions fail to relax, and the system fails to ``gaussify''. This conclusion is special to $1$ dimension; in $d > 1$ dimensions, the same type of system \emph{will} relax to a gaussian state. It is important to keep in mind, however, that the observables for massless particles are typically not correlation functions of the fields themselves, but rather correlation functions of \emph{vertex operators} (exponentials of the fields) or of derivatives of the fields, so that the precise arguments and decay rates for these are slightly different. We will not delve into these details here.

Many properties of (seemingly diverse) gapless systems in one spatial dimension can be obtained within the unifying framework of Luttinger liquid theory \cite{Haldane1981}, which, in its simplest incarnation, can be formulated as a theory of non-interacting massless bosonic fields. However, this formulation relies on linearization of the single-particle dispersion relation, and while this is innocuous for most static properties, it is clearly dangerous when considering relaxation behavior: even a slight dispersion nonlinearity will cause a crossover from non-dispersive to dispersive spreading of the propagator at late enough times, and hence lead to the relaxation that is absent in the free massless bosonic field theory. Thus, any consistent description of the quench dynamics of a one-dimensional system (even an exactly integrable one) using Luttinger liquid theory \emph{must} account for dispersion nonlinearities \cite{Imambekov2012,Seabra2014}, unless the initial state is itself gaussian in terms of the bosonic fields \cite{Cazalilla2016}.

These points were emphasized by Ngo Dinh et al.~\cite{NgoDinh2013}, and also by Sotiriadis \cite{Sotiriadis2017} in follow-up work to \cite{Sotiriadis2016}. These references contain a comprehensive analysis of relaxation in the Luttinger model, and conclude that any weak nonlinearity of the dispersion would ultimately lead to gaussification, in agreement with the intuitive argument sketched above. We refer the reader to these works for a detailed discussion of most of the issues mentioned in this subsection, and to Ref.~\cite{Cazalilla2016} for a general pedagogical discussion of quenches in the Luttinger model.

\subsection{Bosons with pairing}\label{sec:bosons}

We now comment on how our ``gaussification'' results are modified in the case of bosons with pairing. As noted in Section~\ref{sec:propagator}, the propagator for bosons has the general form
\eq{
G(t) = \mathcal{T} e^{-i \int_0^t M(t') \dd{t'}} ,
}
where
\eq{
M(t) = \mat{h(t) & \Delta(t) \\ - \Delta^*(t) & -h^*(t)} ,
}
and where $h^{\dagger} = h$ and $\Delta^T = \Delta$. When $\Delta \neq 0$, the propagator is not unitary, but rather \emph{pseudo-unitary}; it satisfies $G^{\dagger} \eta \, G = \eta$, where $\eta = I_N \oplus - I_N$ and $I_N$ is the $N \times N$ identity matrix. Consequently, the right hand side of Eq.~(\ref{eq:G_sum}) is no longer simply 1, but rather some function of time:
\eq{\label{eq:G_sum_bosons}
\sum_{b=\pm} \sum_y \abs*{G_{xy}^{ab}(t)}^2 = g_x^a(t) > 0 .
}
The non-negligible matrix elements of $G^{ab}_{xy}(t)$ thus have typical magnitude $\sim [\mathcal{V}_x^a(t)]^{-1/2} [g_x^a(t)]^{1/2}$. Repeating the phase-space arguments of Section~\ref{sec:n_function_decay}, we obtain the appropriately modified form of Eq.~(\ref{eq:n_function_decay}):
\eq{\label{eq:n_function_decay_bosons}
\cev{\hat{\psi}^{a_1}_{x_1}(t_1) \hat{\psi}^{a_2}_{x_2}(t_2) \cdots \hat{\psi}^{a_n}_{x_n}(t_n)}
\sim \frac{[g(t)]^{n/2}}{[\mathcal{V}(t)]^{n/2-1}} , 
\vphantom{\bigg|}
}
where
\eq{
g(t) = \big[ g_{x_1}^{a_1}(t_1) g_{x_2}^{a_2}(t_2) \cdots g_{x_n}^{a_n}(t_n) \big]^{1/n} .
}
In general, $g(t)$ could be a complicated function of time, whose form is difficult to predict without some further knowledge of $M(t)$. However, if the hamiltonian is time-independent, we can easily derive the bound (see Appendix~\ref{app:bounds} for details)
\eq{\label{eq:gt_bound_bosons}
1 \leq g(t) \leq \left( \frac{\omega_{\text{max}}}{\epsilon_{\text{min}}} \right)^2 ,
}
where $\omega_{\text{max}}$ is the largest boson mode energy, and $\epsilon_{\text{min}} > 0$ is the smallest eigenvalue of the hermitian matrix $\mathcal{H}$ that defines the hamiltonian via Eq.~(\ref{eq:matrix_H}) (recall that, for bosons, we require $\mathcal{H}$ to be positive-definite). In this case, although the relaxation behavior described by Eq.~(\ref{eq:n_function_decay_bosons}) is complicated, it has a power-law envelope determined entirely by $\mathcal{V}(t)$. 

Time-evolution in bosonic systems approximately described by unstable or metastable quadratic hamiltonians (those whose mode spectra are \emph{not} positive-definite) has been studied in Ref.~\cite{Hackl2018}.

\section{Equilibration to the GGE}\label{sec:equilibration_to_gge}

We have shown in Section~\ref{sec:gaussification} that, if the initial state has the cluster decomposition property (\ref{eq:cluster_decomposition_def}), and if the dynamics are  delocalizing in the sense of Eq.~(\ref{eq:delocalizing_def}), then, as $t \to \infty$, all local $(n > 2)$-point \emph{connected} correlation functions relax to zero in a manner given by Eq.~(\ref{eq:n_function_decay}). Thus, as $t \to \infty$, local correlation functions themselves Wick factorize and are determined entirely by the local 2-point function $\ev{\hat{\psi}_x^a(t) \hat{\psi}_y^b(t)}$, up to corrections of order $1/\mathcal{V}(t)$. All the results of Section~\ref{sec:gaussification} hold for general time-dependent quadratic hamiltonians $\hat{H}(t)$.

If $\hat{H}$ is time-independent, we can go further---as we do now---and show that the system locally equilibrates to the appropriate GGE. In Sections~\ref{sec:conserved_quantities} and \ref{sec:gge_rho}, we construct the GGE density operator and show that it is gaussian; these sections generalize and complete the discussion in Section~\ref{sec:nn_gge}. In Section~\ref{sec:2pt_relaxation}, which generalizes Secton~\ref{sec:nn_2pt_relaxation}, we study equilibration of the local 2-point function to its GGE value. Combined with the results summarized in the previous paragraph, this analysis proves equilibration to the GGE for a wide class of quadratic lattice models, and also furnishes predictions for the leading time-dependence of local observables as $t \to \infty$. A similar analysis is carried out for time-periodic $\hat{H}(t)$ in Section~\ref{sec:floquet_gge}.

\subsection{Conserved quantities}\label{sec:conserved_quantities}

Consider any quadratic time-independent hamiltonian $\hat{H}$ which gives rise to delocalizing dynamics. We begin by showing that in this case all local conserved quantities $\hat{I}_m$ are themselves quadratic in the particle creation and annihilation operators.

By definition of the conserved quantities, we must have $\hat{I}_m(t) = \hat{I}_m(0)$. Without loss of generality, we can take $\hat{I}_m$ to have a definite order $n$ in terms of creation and annihilation operators, because the latter evolve linearly:
\eq{
\hat{I}_m = \sum_{\{x_j\}} \sum_{\{a_j = \pm\}} \mathcal{I}^{a_1 a_2 \cdots a_n}_{x_1 x_2 \cdots x_n} \, 
\hat{\psi}^{a_1}_{x_1} \hat{\psi}^{a_2}_{x_2} \cdots \hat{\psi}^{a_n}_{x_n} .
}
Locality (recall that this means that the $\hat{I}_m$ are \emph{sums} of local densities) requires that the coefficients $\mathcal{I}^{a_1 \cdots a_n}_{x_1 \cdots x_n}$ vanish unless all $\abs{x_i - x_j} \ll L$. Using Eq.~(\ref{eq:Psi(t)}), we have
\eqal{
\hat{I}_m(t) = \sum_{\{x_j, y_j \}} \sum_{\{a_j, b_j = \pm\}} \Big[ 
&\mathcal{I}^{a_1 \cdots a_n}_{x_1 \cdots x_n}
G^{a_1 b_1}_{x_1 y_1}(t) \cdots G^{a_n b_n}_{x_n y_n}(t) \nonumber \\*[-0.5em]
&\times \hat{\psi}^{b_1}_{y_1}  \hat{\psi}^{b_2}_{y_2} \cdots \hat{\psi}^{b_n}_{y_n} \Big] .
}
The conservation condition $\hat{I}_m(0) = \hat{I}_m(t)$ then requires that
\eq{\label{eq:I_m(t)}
\mathcal{I}^{b_1 \cdots b_n}_{y_1 \cdots y_n} = \!
\sum_{\{x_j\}} \! \sum_{\{a_j = \pm\}} \! \mathcal{I}^{a_1 \cdots a_n}_{x_1 \cdots x_n} \,
G^{a_1 b_1}_{x_1 y_1}(t) \cdots G^{a_n b_n}_{x_n y_n}(t) .
}
The same ``phase space'' arguments that we used in the previous section to show decay of all local connected $(n \geq 3)$-point functions also apply to the right hand side of Eq.~(\ref{eq:I_m(t)}); locality of the coefficients $\mathcal{I}^{a_1 \cdots a_n}_{x_1 \cdots x_n}$ here plays the role of cluster decomposition. We conclude that for $n \geq 3$, the right hand side of Eq.~(\ref{eq:I_m(t)}) must vanish as $t \to \infty$ if the dynamics are delocalizing. The left hand side, however, is obviously time-independent and finite. This contradiction proves the claim.

Thus, all local conserved quantities of $\hat{H}$ must be of the form
\eq{
\hat{I}_m = \tfrac{1}{2} \hat{\Psi}^{\dagger} \mathcal{I}_m \hat{\Psi} ,
}
where $\mathcal{I}_m$ is a canonical hermitian $2N \times 2N$ matrix in which each block is banded to ensure locality.

\subsection{GGE density operator}\label{sec:gge_rho}

A quadratic hamiltonian $\hat{H} = \frac{1}{2} \hat{\Psi}^{\dagger} \mathcal{H} \hat{\Psi}$ (with $\mathcal{H}$ positive definite in the case of bosons) can always be diagonalized by a Bogolyubov transformation \cite{VanHemmen1980}; we can introduce new canonical ``quasiparticle'' operators $\{ \hat{\gamma}^{\pm}_n \}$ that obey the same (anti)commutation relations as the $\{ \hat{\psi}^{\pm}_x \}$, and are related to the latter by a linear transformation,
\eq{
\hat{\Psi} = S \, \hat{\Gamma} ,
}
where
\eq{
\hat{\Gamma} = (\hat{\gamma}^-_1, \hat{\gamma}^-_2, \cdots, \hat{\gamma}^-_N, \hat{\gamma}^+_1, \hat{\gamma}^+_2, \cdots, \hat{\gamma}^+_N)^T .
}
The transformation $S$ has the block form
\eq{
S = \mat{U & V^* \\ V & U^*}
}
(to preserve adjoints), satisfies $S^{\dagger} S = I$ for fermions, or $S^{\dagger} \eta S = \eta$ for bosons, where $\eta = I_N \oplus - I_N$ (to preserve the operator algebra), and is diagonalizing: 
\eq{
S^{\dagger} \mathcal{H} S = \Omega \coloneqq \text{diag}(\omega_1, \dots, \omega_N, - \omega_1, \dots, - \omega_N)
}
for fermions, or
\eq{
S^{\dagger} \mathcal{H} S = \eta \, \Omega
} 
for bosons. In terms of the quasiparticle operators, we have
\eq{
\hat{H} = E_0 + \sum_{j=1}^N \omega_j \hat{n}_j ,
}
where
\eq{
\hat{n}_j = \hat{\gamma}^+_j \hat{\gamma}^-_j
}
and $\omega_j \geq 0$ (in the case of bosons, $\omega_j > 0$ is required for physical stability). 

The mode occupation number operators $\hat{n}_j$ commute with $\hat{H}$ and with one another. If the spectrum $\{ \omega_j \}$ is nondegenerate (that is, if $\omega_i = \omega_j$ implies $i = j$), then the set of operators $\{ \hat{n}_j \}$ is uniquely defined, and forms a linear basis for the set of all quadratic conserved quantities of $\hat{H}$. We have already shown (in the previous section) that all local conserved quantities $\hat{I}_m$ of $\hat{H}$ are quadratic if $\hat{H}$ gives rise to delocalizing dynamics. Therefore, in this case we may conclude that the GGE density operator has the form
\eqsub{
\label{eq:gge_rho_general}
\eqal{
\hat{\rho}_{\text{GGE}} 
&= \frac{1}{Z_{\text{GGE}}} \exp( - \sum_m \lambda_m \hat{I}_m ) 
\label{eq:gge_rho_general_a} \\*
&= \frac{1}{Z_{\text{GGE}}} \exp\!\Bigg( - \sum_j \mu_j \hat{n}_j \Bigg) ,
\label{eq:gge_rho_general_b}
}
}
where only the Lagrange multipliers $\{ \lambda_m \}$, or equivalently $\{ \mu_j \}$, are left to be determined by the initial state. We emphasize again that this conclusion relies on two assumptions in addition to $\hat{H}$ being quadratic: (i) that the dynamics are delocalizing and (ii) that the mode spectrum $\{ \omega_j \}$ is nondegenerate.

If the mode spectrum is degenerate, on the other hand, there is some freedom in the choice of diagonalizing canonical transformation $S$, and consequently in the mode operators and conserved quantities. For instance, if $\omega_1 = \omega_2$, consider the family of quasiparticle operators defined by
\eq{
\mat{ \hat{\alpha}^-_1 \\ \hat{\alpha}^-_2 } = Q \mat{ \hat{\gamma}^-_1 \\ \hat{\gamma}^-_2 } ,
}
where $Q \in U(2)$ is any $2 \times 2$ unitary matrix. It is clear that the new number operators
\eq{
\hat{n}_1' = \hat{\alpha}^+_1 \hat{\alpha}^-_1 , \quad \
\hat{n}_2' = \hat{\alpha}^+_2 \hat{\alpha}^-_2
}
also commute with the hamiltonian $\hat{H}$. However, they \emph{do not} in general commute with the old $\hat{n}_1$, $\hat{n}_2$ operators:
\eq{
\comm{\hat{H}}{\hat{n}'_j} = 0
\quad \ \text{but} \quad
\comm{\hat{n}_i}{\hat{n}'_j} \neq 0
\quad \ (i,j = 1,2) .
}
Therefore, the primed ($\hat{n}'$) and unprimed ($\hat{n}$) operators yield \emph{inequivalent} sets of conserved quantities. This ambiguity is fundamental \cite{Fagotti2014}---it is present whenever the mode spectrum is degenerate---and it leads, in principle, to additional dependence on the initial state, as we now describe.

Each inequivalent set of conserved quantities gives rise to its own family of GGE density operators (parameterized by the Lagrange multipliers of that set of quantities). Given an initial state $\hat{\rho}_0$, we must chose the canonical transformation $S$ to also diagonalize the correlations within each degenerate subspace; that is, we must choose $S$ so that, for all pairs $i \neq j$ such that $\omega_i = \omega_j$, we have
\eq{
\ev{\hat{\gamma}^+_i \hat{\gamma}^-_j} \coloneqq \Tr(\hat{\gamma}^+_i \hat{\gamma}^-_j \, \hat{\rho}_0) = 0 .
}
In the case of fermions, we must also choose $S$ to ensure that, whenever $\omega_i = \omega_j = 0$,
\eq{
\ev{\hat{\gamma}^+_i \hat{\gamma}^+_j} = 0 .
}
It is always possible to find a canonical transformation $S$ that diagonalizes $\hat{H}$ and also satisfies these conditions. The GGE density operator can then be constructed using the associated mode operators in the usual manner, following Eq.~(\ref{eq:gge_rho_general_b}).

Thus in general $\hat{\rho}_{\text{GGE}}$, written in the form (\ref{eq:gge_rho_general_b}), depends on the initial state $\hat{\rho}_0$ in two distinct ways: (i) the definition of the occupation numbers operators $\{ \hat{n}_j \}$ corresponding to degenerate modes $\{ \omega_j \}$ of $\hat{H}$, and (ii) the values of the Lagrange multipliers $\{ \mu_j \}$. 

The general construction of the GGE density operator that we have outlined in this section can be applied to any quadratic hamiltonian $\hat{H}$ that gives rise to delocalizing dynamics; it will indeed yield a density operator $\hat{\rho}_{\text{GGE}}$ that correctly describes all local observables of the system at late times (as we demonstrate in the next section). However, we have in some sense ``cheated'' by phrasing our general construction in terms of the mode occupation numbers $\{ \hat{n}_j \}$ rather than in terms of the local conserved quantities $\{ \hat{I}_m \}$. Since we are studying \emph{local} relaxation, the latter are really the quantities of fundamental importance. 

From a more fundamental point of view, then, a set $\{ \hat{n}_j \}$ is \emph{admissible} only if, by taking linear combinations of the $\hat{n}_j$, one can construct a maximal set of local conserved quantities $\{ \hat{I}_m \}$ (recall that the set $\{ \hat{I}_m \}$ is maximal if any local conserved quantity $\hat{I}$ that commutes with all of the $\hat{I}_m$ can be expressed as a linear combination of them). Given a maximal set $\{ \hat{I}_m \}$, we can always obtain a corresponding admissible set $\{ \hat{n}_j \}$ by finding the Bogolyubov transformation $S$ that simultaneously diagonalizes the $\hat{I}_m$. Sets $\{ \hat{n}_j \}$ that are \emph{inadmissible} can---regardless of the initial state---be ignored for the purpose of writing down $\hat{\rho}_{\text{GGE}}$, and one only needs to use initial correlations to distinguish between admissible sets. Thus, the construction outlined in this section, although valid, might overestimate the degree to which the GGE depends on the initial state.

\subsection{Relaxation of the local 2-point function}\label{sec:2pt_relaxation}

Having constructed the GGE density operator, we now study relaxation towards it by analyzing the long-time behavior of the local 2-point function. As in the example of Section~\ref{sec:nn_chain}, this part of the analysis requires us to make an additional assumption about the initial state $\hat{\rho}_0$; roughly speaking, we need to exclude situations in which the initial profiles of local conserved densities are inhomogeneous on length scales comparable to the system size. True local equilibration in such cases occurs on timescales of the order of the linear dimension $L$ of the system, simply because that is how long it takes a locally conserved density to flow across the system.

In order to formulate this assumption precisely, recall that the local conserved quantities are of the form
\eq{
\hat{I}_m = \sum_x \hat{\mathcal{I}}_{m,x} ,
}
where the density $\hat{\mathcal{I}}_{m,x}$ is supported in a finite region centered at position $x$. Define the ``local excess density''
\eq{
\delta \mathcal{I}_m(x_0; r) \coloneqq \frac{1}{\text{Vol}(B_r)} \! \sum_{x \in B_r(x_0)} \! \ev{\hat{\mathcal{I}}_{m,x}}
\, - \frac{\ev{\hat{I}_m}}{\text{Vol(Sys)}} ,
}
where $B_r(x_0)$ is the $d$-dimensional ball of radius $r$ centered at $x_0$, $\text{Vol}(B_r)$ is the volume of this ball, and $\text{Vol(Sys)}$ is the volume of the entire system.

We assume that these excess densities can be made small by taking $r$ sufficiently large (but keeping $r$ fixed as $\text{Vol(Sys)} \to \infty$):
\eqal{\label{eq:gen_homogeneous}
\exists \, r : \, \delta \mathcal{I}_m(x_0; r) = O(r^{-d}) \ \ &\forall \ x_0, m \nonumber \\*
&\text{as} \ \ \text{Vol(Sys)} \to \infty .
}

We emphasize that the ``gaussification'' results of Section~\ref{sec:gaussification} hold even when this assumption is violated (the results of Sections~\ref{sec:conserved_quantities} and \ref{sec:gge_rho} hold as well). Thus, if the initial state violates Eq.~(\ref{eq:gen_homogeneous}), the natural description of the local state of the system at late times is in terms of a time-dependent gaussian density matrix, of the form
\eq{
\hat{\rho}_1(t) = \frac{1}{Z_1(t)} \exp( - \tfrac{1}{2} \hat{\Psi}^{\dagger} K(t) \hat{\Psi} ) , 
}
where $K(t)$ is a canonical hermitian matrix (it satisfies Eq.~(\ref{eq:canonical_hermitian_def})) that must be chosen so that
\eq{
\Tr( \hat{\psi}^a_x(t) \hat{\psi}^b_y(t) \, \hat{\rho}_1(t) ) = \ev{ \hat{\psi}^a_x(t) \hat{\psi}^b_y(t) }
} 
for all pairs of indices $a,b = \pm$ and positions $x,y$ with $\abs{x-y}$ finite in the limit of infinite system size.

Also as in the example of Section~\ref{sec:nn_chain}, this part of the analysis requires more detailed knowledge of the spectrum of the hamiltonian, or equivalently, of the propagator, than is needed to show ``gaussification''. Consequently, our treatment will be somewhat schematic.

In terms of the matrix $S$ of the Bogolyubov transformation $\hat{\Psi} = S \, \hat{\Gamma}$ that diagonalizes the hamiltonian $\hat{H}$, the propagator may be written as
\eq{\label{eq:gen_Gt}
G(t) = S e^{-i \Omega t} \, S^{-1} ,
}
where
\eq{
\Omega \coloneqq \text{diag}(\omega_1, \omega_2, \dots, \omega_N, - \omega_1, - \omega_2, \dots, - \omega_N) ,
}
and $\{ \omega_j \geq 0 \}$ is the spectrum of quasiparticle excitations. This form of $G(t)$ is valid for both fermions and for bosons; the difference between the two is the unitarity or pseudo-unitarity of the matrix $S$. It is standard to regard the $2N$ columns of $S$ as eigenvectors of a fictitious single-particle problem whose eigenvalue spectrum is symmetric about zero (while keeping in mind that, for bosons, the eigenvectors are orthonormal with respect to $\eta = I_N \oplus - I_N$ rather than $I_{2N}$). If we label these eigenvectors by their energy $\varepsilon$ ($\varepsilon = \pm \omega_j$) and additional quantum numbers $\sigma$, so that $(\varepsilon,\sigma)$ together form a complete set, we can write
\eq{
G_{xy}^{ab}(t) = \sum_{\varepsilon, \sigma} S_x^a(\varepsilon,\sigma) \, e^{-i \varepsilon t} \, (S^{-1})_y^b(\varepsilon,\sigma) .
}
The equal-time 2-point function is then given by
\eqal{\label{eq:gen_2pt}
\ev{\hat{\psi}_x^{-a}(t) \hat{\psi}_y^b(t)}
= \sum_{\varepsilon, \sigma, \, \varepsilon', \sigma'} \Big\{
&[S_x^a(\varepsilon,\sigma)]^* S_y^b(\varepsilon',\sigma') \, e^{-i (\varepsilon - \varepsilon') t} \nonumber \\*[-0.5em]
&\times F(\varepsilon,\sigma;\varepsilon',\sigma') \Big\} ,
}
where
\eq{\label{eq:gen_F}
F(\varepsilon,\sigma;\varepsilon',\sigma') \coloneqq
\ev{\hat{\Gamma}^{\dagger}(\varepsilon,\sigma) \hat{\Gamma}(\varepsilon',\sigma')} ,
}
and where
\eq{\label{eq:gen_Gamma}
\hat{\Gamma}(\varepsilon,\sigma) \coloneqq \begin{cases}
\hat{\gamma}(\varepsilon,\sigma) &\text{if} \quad \varepsilon \geq 0 \\
\hat{\gamma}^{\dagger}(-\varepsilon,\sigma) &\text{if} \quad \varepsilon < 0
\end{cases} .
}
The GGE value of the same 2-point function is
\eq{\label{eq:gen_2pt_gge}
\ev{\hat{\psi}_x^{-a} \hat{\psi}_y^b}_{\text{GGE}}
= \sum_{\varepsilon, \sigma}
[S_x^a(\varepsilon,\sigma)]^* S_y^b(\varepsilon,\sigma) \, 
\ev{\hat{\Gamma}^{\dagger}(\varepsilon,\sigma) \hat{\Gamma}(\varepsilon,\sigma)} .
}
Equations~(\ref{eq:gen_2pt}--\ref{eq:gen_2pt_gge}) are the obvious generalizations of Eqs.~(\ref{eq:nn_2pt}), (\ref{eq:nn_F}) and (\ref{eq:nn_2pt_gge}).

In the limit of large system size, the spectrum $\{ \varepsilon \}$ will in general consist of a continuous part due to spatially extended quasiparticle states and a discrete part due to localized states. For now, we assume that all quasiparticle states are extended. We will discuss what happens when the spectrum includes a discrete part coming from localized states in Section~\ref{sec:localized_states} (see also the comments in Section~\ref{sec:propagator}).

Since the spectrum is by assumption purely continuous in the limit of large system size, the sums over $\varepsilon$ and $\varepsilon'$ in Eq.~(\ref{eq:gen_2pt}) become integrals in this limit (whether the other quantum numbers $\sigma$ are discrete or continuous is less important). The $t \to \infty$ asymptotics of the $(\varepsilon, \varepsilon')$-integral is then determined by the analytic structure of the function $F$. This structure can in turn be deduced from general arguments of the type used in Section~\ref{sec:nn_chain}. Inverting the Bogoliubov transformation, we have
\eqal{\label{eq:gen_F_exp}
F(\varepsilon,\sigma;\varepsilon',\sigma')
= \sum_{a,b = \pm} \sum_{x,y} \Big\{
&[(S^{-1})_x^a(\varepsilon,\sigma)]^* (S^{-1})_y^b(\varepsilon',\sigma') \nonumber \\*[-0.5em]
&\times \ev{\hat{\psi}_x^{-a} \hat{\psi}_y^b} \Big\} .
}
The sums over $x$ and $y$ in Eq.~(\ref{eq:gen_F_exp}) may be performed with respect to the central coordinate $(x + y)/2$ and relative coordinate $(x - y)$. The sum over the relative coordinate converges absolutely (due to clustering of correlations), whereas the central coordinate is summed over the whole system (because the states are extended). It follows that $F$ can become singular only along ``curves'' in $(\varepsilon,\sigma;\varepsilon',\sigma')$-space, which we may identify with the zero sets of appropriate functions $\mathcal{C}_j(\varepsilon,\sigma;\varepsilon',\sigma')$. The most obvious such curve is the trivial one, $(\varepsilon',\sigma') = (\varepsilon,\sigma)$, which may be identified with the function $\mathcal{C}_0(\varepsilon,\sigma;\varepsilon',\sigma') \sim (\varepsilon - \varepsilon')(\sigma - \sigma')$; additional curves $\mathcal{C}_j$ can occur if the initial state has an appropriate order (in Section~\ref{sec:nn_chain}, for instance, we found that such curves were present if the initial state had a density wave with nonzero wavevector $q$). We conclude that $F$ has the (highly schematic) general form
\eqal{\label{eq:gen_F_structure}
F(\varepsilon, \sigma;  \varepsilon', \sigma') 
&= \delta(\varepsilon - \varepsilon') \delta(\sigma - \sigma') \, \ev{\hat{\Gamma}^{\dagger}(\varepsilon,\sigma) \hat{\Gamma}(\varepsilon,\sigma)} \nonumber \\*[0.4em]
&\quad + \sum_j \delta(\mathcal{C}_j(\varepsilon, \sigma; \varepsilon',\sigma')) \, f_j(\varepsilon, \sigma) \nonumber \\*
&\quad + f(\varepsilon, \sigma;  \varepsilon', \sigma') ,
}
where the sum in the second line is over a finite number of curves $\mathcal{C}_j$ that, as a consequence of our assumption (\ref{eq:gen_homogeneous}), remain distinct from the trivial curve in the limit of infinite system size. The various deltas functions represent all of the possible singular dependence of $F$ on its arguments; $\ev{\hat{\Gamma}^{\dagger}(\varepsilon,\sigma) \hat{\Gamma}(\varepsilon,\sigma)}$, $f_j(\varepsilon, \sigma)$ and $f(\varepsilon, \sigma;  \varepsilon', \sigma')$ are smooth functions in the relevant domains of integration.

Taking account of this structure, Eq.~(\ref{eq:gen_2pt}) becomes
\eqal{\label{eq:gen_2pt_structure}
\ev{\hat{\psi}_x^{-a}(t) \hat{\psi}_y^b(t)}
= &\int_{\varepsilon, \sigma}
[S_x^a(\varepsilon,\sigma)]^* S_y^b(\varepsilon,\sigma) \,
\ev{\hat{\Gamma}^{\dagger}(\varepsilon,\sigma) \hat{\Gamma}(\varepsilon,\sigma)} \nonumber \\*
&+ \sum_j [\delta C_j(t)]_{xy}^{ab} \, + [\delta C(t)]_{xy}^{ab} \vphantom{\int} .
}
The first term reproduces the GGE result, Eq.~(\ref{eq:gen_2pt_gge}). The remaining $\delta C_j$ and $\delta C$ pieces come from the second and third terms in Eq.~(\ref{eq:gen_F_structure}) respectively. 

Let us first analyze the $\delta C$ term,
\eqal{\label{eq:gen_dCt_int}
[\delta C(t)]_{xy}^{ab} = \int_{\varepsilon, \varepsilon'} \int_{\sigma, \sigma'} \Big\{
&[S_x^a(\varepsilon,\sigma)]^* \, S_y^b(\varepsilon',\sigma') \, e^{-i (\varepsilon - \varepsilon') t} \nonumber \\*[-0.3em]
&\times f(\varepsilon,\sigma;\varepsilon',\sigma') \Big\} .
}
The behavior of the integral as $t \to \infty$ can be extracted from a straightforward stationary phase analysis (apart from the factor $e^{-i (\varepsilon - \varepsilon') t}$, the integrand is a smooth function of the integration variables). The phase function $\varphi(\varepsilon,\varepsilon') = (\varepsilon - \varepsilon')$ clearly lacks stationary points, so the dominant contribution to the integral as $t \to \infty$ comes from the \emph{corners} of the $(\varepsilon, \varepsilon')$-integration region. Near each corner, the smooth function $f$ can be regarded as a function of $\sigma$ and $\sigma'$ alone. The integrals over $\sigma$ and $\sigma'$ will then yield factors proportional to the (local) density of states $g(\varepsilon)$ and $g(\varepsilon')$ near the band edges. We are led to conclude that, as $t \to \infty$,
\eq{
\delta C(t) \sim (\#) \times \abs{\int \dd{\varepsilon} g(\varepsilon) e^{-i \varepsilon t}}^2 .
}
If, as is often the case, the density of states near the band edge has the form
\eq{
g(\varepsilon) \sim \varepsilon^s ,
}
then $\int \dd{\varepsilon} g(\varepsilon) \, e^{-i \varepsilon t} \sim t^{-(1+s)} \int \dd{z} z^s e^{-i z}$, and we obtain the estimate
\eq{\label{eq:gen_dCt}
\delta C(t) \sim t^{-2(1+s)} .
}
This result assumes that $f$ does not vanish at the corners of the $(\varepsilon, \varepsilon')$-integration region.
Generically, this will be the case.
In special fine-tuned circumstances, in which $f$ does vanish at the corners, the exponent of the power law may be larger (more negative).

We can perform a similar stationary phase analysis of each $\delta C_j$ term in Eq.~(\ref{eq:gen_2pt_structure}). In this case, the phase function $\varphi$ is the restriction of $(\varepsilon - \varepsilon')$ to the curve $\mathcal{C}_j$. If $\varphi$ is nonstationary along this curve, and if the curve terminates at the boundary of the $(\varepsilon, \varepsilon')$ integration region, then the same reasoning that we applied to $\delta C$ in the previous paragraph yields the estimate
\eq{\label{eq:gen_dCjt}
\delta C_j(t) \sim t^{-(1+s)} .
}
Again, this result may be modified if the initial state or final hamiltonian are fine-tuned.
More complicated time-dependence will occur if the phase function $\varphi$ is stationary somewhere along the curve $\mathcal{C}_j$; such a contribution, if present, will likely dominate the $t \to \infty$ relaxation behavior. However, this must be analyzed on a case-by-case basis, and we will not attempt to make any further statements about the general case.

If the hamiltonian $\hat{H}$ is translation-invariant, then one typically has
\eq{
g(\varepsilon) \sim \varepsilon^{(d/2 - 1)}
}
at each band edge, where $d$ is the dimension of space. In this case the above estimates become
\eqsub{
\label{eq:trans_inv_dCt}
\eqal{
\delta C(t) &\sim t^{-d} \, , 
\label{eq:trans_inv_dCt_a} \\
\delta C_j(t) &\sim t^{-d/2} .
\label{eq:trans_inv_dCt_b}
}
}
The results obtained in the example of Section~\ref{sec:nn_chain}---Eqs.~(\ref{eq:nn_dCjt}) and (\ref{eq:nn_dCt})---are recovered if one sets $d=1$. 

It is interesting to compare Eq.~(\ref{eq:trans_inv_dCt}), which gives the asymptotic relaxation of the 2-point function in a translation-invariant lattice system, to the asymptotic power law with which such a system should gaussify according to the results of Section~\ref{sec:gaussification}. The latter power law is set by the lowest nonvanishing $(n>2)$-point connected correlation function. Assuming that this is $n=4$, Eq.~(\ref{eq:n_function_decay}) suggests that the system gaussifies like $\sim [\mathcal{V}(t)]^{-1}$, where $\mathcal{V}(t)$ is the volume on which the 1-particle propagator is meaningfully supported. In a translation-invariant lattice model, the propagator spreads at the maximal group velocity, so we expect this volume to grow like $\mathcal{V}(t) \sim t^d$. Hence we conclude that the system gaussifies like $\sim t^{-d}$. If there is a density wave of one or more of the conserved quantities in the initial state, then $\delta C_j(t)$ terms are present in the 2-point function; these relax like $\sim t^{-d/2}$ by Eq.~(\ref{eq:trans_inv_dCt_b}). Thus the system \emph{first gaussifies like $\sim t^{-d}$, and then relaxes to the GGE like $\sim t^{-d/2}$}. If, on the other hand, the initial state lacks such order, then only the $\delta C(t)$ term is present in the 2-point function; this relaxes like $\sim t^{-d}$ by Eq.~(\ref{eq:trans_inv_dCt_a}), so \emph{gaussification and relaxation to the GGE both occur with the power law $\sim t^{-d}$}.

Notice that gaussification and relaxation of the 2-point function are controlled (in translation-invariant systems) by fundamentally different aspects of the band structure:
gaussification is controlled by the maximal group velocity---typically a property of the middle of the band(s)---whereas relaxation of the 2-point function is controlled by the density of single-particle levels at the band edge(s).

\section{Effects due to localized states}\label{sec:localized_states}

Assume now that the quasiparticle spectrum of $\hat{H}$ contains, in the limit of large system size, both discrete localized states and a continuum of extended states. We can write the diagonalizing Bogoliubov transformation as
\eq{\label{eq:bogoliubov_ext_loc}
\hat{\psi}_x^a = \int_{\varepsilon} \int_{\sigma} S_x^a(\varepsilon, \sigma) \hat{\Gamma}(\varepsilon, \sigma)
+ \sum_{b = \pm} \sum_j R_{xj}^{ab} \, \hat{\gamma}_j^b ,
}
where, as before,
\eq{
\hat{\Gamma}(\varepsilon,\sigma) \coloneqq \begin{cases}
\hat{\gamma}(\varepsilon,\sigma) &\text{if} \quad \varepsilon \geq 0 \\
\hat{\gamma}^{\dagger}(-\varepsilon,\sigma) &\text{if} \quad \varepsilon < 0
\end{cases} .
}
The operator $\hat{\gamma}^{\dagger}(\omega,\sigma)$ creates a quasiparticle in the continuum level with energy $\omega \geq 0$ and additional quantum numbers $\sigma$; the operator $\hat{\gamma}_j^+$ creates a quasiparticle in the discrete level $j$ with energy $\omega_j \geq 0$.

\subsection{The propagator}\label{sec:loc_propagator}

The propagator splits naturally into two pieces:
\eq{\label{eq:loc_ext_Gt}
G(t) = G_{\text{ext}}(t) + G_{\text{loc}}(t) ,
}
where the first piece $G_{\text{ext}}(t)$ involves only the extended states, and the second piece $G_{\text{loc}}(t)$ involves only the localized states. For fermions,
\eq{\label{eq:ext_Gt}
[G_{\text{ext}}(t)]_{xy}^{ab} = \int_{\varepsilon} \int_{\sigma} S_x^a(\varepsilon, \sigma) \, e^{-i \varepsilon t} \, [S_y^b(\varepsilon, \sigma)]^*
}
and
\eq{\label{eq:loc_Gt}
[G_{\text{loc}}(t)]_{xy}^{ab} = \sum_{c = \pm} \sum_j R_{xj}^{ac} \, e^{i c \, \omega_j t} \, [R_{yj}^{bc}]^* .
}
For bosons, one must multiply the integrand in Eq.~(\ref{eq:ext_Gt}) by $-b \, \text{sgn}(\varepsilon)$, and the summand in Eq.~(\ref{eq:loc_Gt}) by $bc$.

The dynamics of the propagator $G(t)$, as defined in Section~\ref{sec:propagator}, are thus in general the sum of a \emph{delocalizing} part, due to $G_{\text{ext}}(t)$, and a \emph{localizing} part, due to $G_{\text{loc}}(t)$. We have already discussed general properties of $G_{\text{ext}}(t)$ in Sections~\ref{sec:propagator} and \ref{sec:rates}. Let us now briefly discuss general properties of $G_{\text{loc}}(t)$:

Each level $j$ of the discrete spectrum is exponentially localized near some position $x_j$; in other words,
\eq{
R_{xj}^{ac} \sim e^{-\abs{x - x_j}/\zeta_j}
\quad \text{for} \quad
\abs{x - x_j} \gtrsim \zeta_j  \vphantom{\sum} ,
}
where $\zeta_j > 0$ is the decay length. It follows from Eq.~(\ref{eq:loc_Gt}) that
\eq{
[G_{\text{loc}}(t)]_{xy}^{ab} \sim e^{-\abs{x-y}/\zeta_x} 
\quad \text{for} \quad
\abs{x-y} \gtrsim \zeta_x ,
}
where $\zeta_x$ is roughly the largest decay length of the states localized near $x$. Thus, $[G_{\text{loc}}(t)]_{xy}^{ab}$ is negligible whenever $\abs{x-y} \gg \zeta_x$. Given a position $x$, we may restrict the sum over $j$ in Eq.~(\ref{eq:loc_Gt}) to those levels that are localized within a few decay lengths $\zeta_j$ of $x$, because the remaining levels give negligible contributions. Finally, for fixed $x$ and $y$, the propagator $[G_{\text{loc}}(t)]_{xy}^{ab}$ oscillates forever without decaying as $t \to \infty$.

\vspace{5pt}

\subsection{Gaussification}\label{sec:loc_gaussification}

Having understood how the propagator is modified, let us study how the localized states affect gaussification and the conclusions of Section~\ref{sec:gaussification}. Consider the time-dependent connected $n$-point function. Equation~(\ref{eq:connected_n_function}) becomes
\begin{widetext}
\eq{\label{eq:connected_n_ext_loc}
\cev{\hat{\psi}^{a_1}_{x_1}(t_1) \cdots \hat{\psi}^{a_n}_{x_n}(t_n)}
= \sum_{\{ y_i \}} \sum_{\{ b_i = \pm \}} \Big\{
[G_{\text{ext}}(t_1) + G_{\text{loc}}(t_1)]^{a_1 b_1}_{x_1 y_1} \cdots 
[G_{\text{ext}}(t_n) + G_{\text{loc}}(t_n)]^{a_n b_n}_{x_n y_n} \, 
\cev{\hat{\psi}^{b_1}_{y_1} \cdots \hat{\psi}^{b_n}_{y_n}} \Big\} .
}
\end{widetext}
Write this as
\eqml{
\cev{\hat{\psi}^{a_1}_{x_1}(t_1) \cdots \hat{\psi}^{a_n}_{x_n}(t_n)} \vphantom{\sum} \\*
= \sum_{k=0}^n \cev{\hat{\psi}^{a_1}_{x_1}(t_1) \cdots \hat{\psi}^{a_n}_{x_n}(t_n)}_{k-\text{loc}} ,
}
where $\cev{\cdots}_{k-\text{loc}}$ contains all terms in Eq.~(\ref{eq:connected_n_ext_loc}) that have $k$ factors of $G_{\text{loc}}$ and $(n-k)$ factors of $G_{\text{ext}}$.

We have already studied the contribution $\cev{\cdots}_{0-\text{loc}}$, in which all the propagators are $G_{\text{ext}}$, in detail in Section~\ref{sec:n_function_decay}. As $t \to \infty$, $\cev{\cdots}_{0-\text{loc}}$ decays to zero as described by Eq.~(\ref{eq:n_function_decay}). Next, consider the contribution $\cev{\cdots}_{1-\text{loc}}$, in which a single propagator is $G_{\text{loc}}$. According to the discussion above, this contribution is significant (at any time $t$) only if one or more of the $x_i$ are located within a few decay lengths of a localized state. The sum over the corresponding $y_i$ is restricted by the propagator $[G_{\text{loc}}(t_i)]_{x_i y_i}^{a_i b_i}$ to a region of volume $\sim \zeta_{x_i}^d$ around $x_i$. Repeating the analysis of Section~\ref{sec:n_function_decay}, the sums over the relative $y$-coordinates converge absolutely due to cluster decomposition, so the entire $y$-sum yields a finite, $t$-independent contribution as $t \to \infty$. Meanwhile, typical matrix elements of the propagators are of order $G_{\text{ext}}(t) \sim [\mathcal{V}(t)]^{-1/2}$ and $G_{\text{loc}}(t) \sim 1/\zeta_{x_i}^d$. We conclude that, as $t \to \infty$,
\eq{\label{eq:1_loc_decay}
\cev{\cdots}_{1-\text{loc}} \sim [\mathcal{V}(t)]^{-(n-1)/2} .
}
This decay is \emph{faster}, by a factor of $[\mathcal{V}(t)]^{-1/2}$, than that of $\cev{\cdots}_{0-\text{loc}}$.

By similar reasoning, we conclude that
\eq{
\cev{\cdots}_{k-\text{loc}} \sim [\mathcal{V}(t)]^{-(n-k)/2} \quad \ (k \geq 1) .
}
Note that, for $k \geq 1$, $\cev{\cdots}_{(k+1)-\text{loc}}$ decays \emph{slower}, by a factor of $[\mathcal{V}(t)]^{-1/2}$, than does $\cev{\cdots}_{k-\text{loc}}$. Thus, the leading $t \to \infty$ behavior of the connected $(n>2)$-point function is
\eqal{
&\cev{\hat{\psi}^{a_1}_{x_1}(t_1) \cdots \hat{\psi}^{a_n}_{x_n}(t_n)} \vphantom{\sum} \nonumber \\*
&\quad \sim \cev{\hat{\psi}^{a_1}_{x_1}(t_1) \cdots \hat{\psi}^{a_n}_{x_n}(t_n)}_{n-\text{loc}}
+ O([\mathcal{V}(t)]^{-1/2}) .
}
Only the fully localized contribution $\cev{\cdots}_{n-\text{loc}}$ survives in the limit $t \to \infty$. Let us analyze this term in more detail:
\eqal{
\cev{\hat{\psi}^{a_1}_{x_1}(t_1) \cdots \hat{\psi}^{a_n}_{x_n}(t_n)}_{n-\text{loc}} \vphantom{\sum} \nonumber \\*
= \sum_{\{ y_i \}} \sum_{\{ b_i = \pm \}} \Big\{
&[G_{\text{loc}}(t_1)]^{a_1 b_1}_{x_1 y_1} \cdots [G_{\text{loc}}(t_n)]^{a_n b_n}_{x_n y_n} \nonumber \\*[-0.7em]
&\times \cev{\hat{\psi}^{b_1}_{y_1} \cdots \hat{\psi}^{b_n}_{y_n}} \Big\} .
}
Using Eq.~(\ref{eq:loc_Gt}) or its bosonic version, we have
\eq{
\sum_{b = \pm} \sum_y \, [G_{\text{loc}}(t)]^{ab}_{xy} \, \hat{\psi}_y^b
= \sum_{c = \pm} \sum_j R_{xj}^{ac} \, e^{i c \, \omega_j t} \, \hat{\gamma}_j^c .
}
Consequently,
\eqal{\label{eq:connected_n-loc}
\cev{\hat{\psi}^{a_1}_{x_1}(t_1) \cdots \hat{\psi}^{a_n}_{x_n}(t_n)}_{n-\text{loc}} \vphantom{\sum} \nonumber \\*
= \sum_{\{ j_i \}} \sum_{\{ c_i = \pm \}} \Big\{
&R_{x_1 j_1}^{a_1 c_1} \cdots R_{x_n j_n}^{a_n c_n} \, 
e^{i \sum_{\ell=1}^n c_{\ell} \omega_{j_{\ell}} t_{\ell}} \nonumber \\*[-0.7em]
&\times \cev{\hat{\gamma}^{c_1}_{j_1} \cdots \hat{\gamma}^{c_n}_{j_n}} \Big\} .
}
Each sum over $j_i$ in Eq.~(\ref{eq:connected_n-loc}) may be restricted to those levels that are localized near $x_i$, in accordance with our previous discussion. It is evident that the localized contribution is negligible as $t \to \infty$ if and only if $\cev{\hat{\gamma}^{c_1}_{j_1} \hat{\gamma}^{c_2}_{j_2} \cdots \hat{\gamma}^{c_n}_{j_n}}$ itself is negligible. In bosonic systems prepared in generic initial states, this condition will be violated as soon as there is a single localized level. This is because the $n$th cumulant of the occupation of this level, $\cev{(\hat{\gamma}^+ \hat{\gamma}^-)^n}$, will be nonzero in general. In a fermionic system, on the other hand, the occupation of a single localized level is characterized entirely by the expectation value $\ev{\hat{\gamma}^+ \hat{\gamma}^-}$, so these higher cumulants all vanish.

Thus, consider a system of fermions in which the quasiparticle spectrum of $\hat{H}$ contains, in addition to a continuum of extended states, a set of discrete levels $\{ j \}$ that are localized near positions $\{ x^*_j \}$ with decay lengths $\{ \zeta_j \}$. Assume that the initial state $\hat{\rho}_0$ has a finite correlation length $\xi$, and that for each pair $(i,j)$ of localized levels, $\abs*{x^*_i - x^*_j} \gg \xi + \zeta_i + \zeta_j$. Then any connected function involving the operators of two distinct levels $i \neq j$, such as $\cev{\hat{\gamma}_i^a \, \hat{\gamma}_j^b \cdots}$, is negligible. Of course, any connected function involving three or more operators of the \emph{same} level, such as $\cev{\hat{\gamma}_j^a \, \hat{\gamma}_j^b \, \hat{\gamma}_j^c \cdots}$, vanishes identically. It follows that all $\cev{\cdots}_{(k>2)-\text{loc}}$ contributions to the connected $n$-point function are negligible. Since the $k=0,2$ terms decay in the same manner with time, and since the $k=1$ term decays faster than either of them, we reach the following somewhat surprising conclusion: 

Discrete localized levels in the quasiparticle spectrum of a quadratic fermion hamiltonian $\hat{H}$ \emph{have a negligible effect on gaussification} if (i) the initial state has a finite correlation length $\xi$, and (ii) the spatial distance between any pair of localized levels is large relative to $\xi$.

\subsection{GGE density operator}\label{sec:loc_gge}

Next, we study how the localized states affect the conclusions of Section~\ref{sec:equilibration_to_gge}. Before considering equilibration, we must revisit the construction of the GGE density operator itself. 

In Section~\ref{sec:conserved_quantities}, we showed that, for any quadratic time-independent hamiltonian $\hat{H}$ which gives rise to delocalizing dynamics, all local conserved charges $\hat{I}_m$ are themselves quadratic in the particle creation and annihilation operators. Let us see how this argument changes with localized states. Equation~(\ref{eq:I_m(t)}) remains valid, but each propagator factor now has an extended piece and a localized piece:
\eqal{\label{eq:I(t)_ext_loc}
\mathcal{I}^{b_1 \cdots b_n}_{y_1 \cdots y_n} =
\sum_{\{ x_i \}} \! \sum_{\{ a_i = \pm \}} \!\! \Big\{
&\mathcal{I}^{a_1 \cdots a_n}_{x_1 \cdots x_n}
[G_{\text{ext}}(t) + G_{\text{loc}}(t)]^{a_1 b_1}_{x_1 y_1} \cdots \nonumber \\*[-0.5em]
&\cdots [G_{\text{ext}}(t) + G_{\text{loc}}(t)]^{a_n b_n}_{x_n y_n} \! \Big\} .
}
As we did with the connected $n$-point function in Section~\ref{sec:loc_gaussification}, write this as
\eq{\label{eq:I(t)_exp}
\mathcal{I}^{b_1 \cdots b_n}_{y_1 \cdots y_n}
= \sum_{k=0}^n \, [\mathcal{I}_{k-\text{loc}}(t)]^{b_1 \cdots b_n}_{y_1 \cdots y_n} ,
}
where $\mathcal{I}_{k-\text{loc}}(t)$ contains all terms in Eq.~(\ref{eq:I(t)_ext_loc}) that have $k$ factors of $G_{\text{loc}}(t)$ and $(n-k)$ factors of $G_{\text{ext}}(t)$. Repeating the arguments of Section~\ref{sec:loc_gaussification}, we conclude that for $n > 2$, only the $\mathcal{I}_{n-\text{loc}}(t)$ contribution survives as $t \to \infty$. Then, Eq.~(\ref{eq:I(t)_exp}) requires that $\mathcal{I}_{n-\text{loc}}$ actually be time-independent, and that $\mathcal{I} = \mathcal{I}_{n-\text{loc}}$. It is clear that the corresponding local conserved quantities are those that can be built from the quasiparticle operators $\hat{\gamma}^{\pm}_j$ of the localized levels:
\eq{
\hat{I}_m \sim \hat{\gamma}_{j_1}^{c_1} \, \hat{\gamma}_{j_2}^{c_2} \, \cdots \hat{\gamma}_{j_n}^{c_n} , 
\quad \ \text{with} \quad
\sum_{i=1}^n c_i \, \omega_{j_i} = 0 .
}
In addition, the participating levels $\{ j_i \}$ must all be localized in the same region of space (otherwise $\hat{I}_m$ will not be local). Conversely, \emph{all} local conserved charges involving products of $n > 3$ creation or annihilation operators must be of this form. Thus, in a bosonic system, the existence of even a single localized level leads to non-quadratic local conserved charges (powers of the occupation of this level, $(\hat{\gamma}^+ \hat{\gamma}^-)^n$). In a fermionic system, however, one can only construct non-quadratic local conserved charges if there are two or more localized levels close enough to one another in space (how close depends on how local we want the charges to be).

We conclude that, in a quadratic bosonic system, the GGE (defined in terms of local conserved quantities) is gaussian if and only if there are no localized levels at all, whereas in a quadratic fermionic system, the GGE remains gaussian to an excellent approximation even when localized levels do exist, as long as they are located sufficiently far apart in space. In the latter case, the mode occupation numbers $\hat{n}_j = \hat{\gamma}^+_j \hat{\gamma}^-_j$ of these levels are local conserved charges, and must be included in $\hat{\rho}_{\text{GGE}}$. The general analysis of Section~\ref{sec:gge_rho} does not require modification.

\subsection{Equilibration to the GGE}\label{sec:loc_equilibration_to_gge}

Finally, let us consider equilibration. As in Section~\ref{sec:2pt_relaxation}, we must make an additional assumption on the initial state, Eq.~(\ref{eq:gen_homogeneous}), to exclude situations in which the initial profiles of local conserved densities are inhomogeneous on length scales comparable to the system size.

Following Section~\ref{sec:loc_gaussification}, we may identify three contributions to the equal-time 2-point function:
\eq{\label{eq:loc_ext_2pt}
\ev{\hat{\psi}_x^{-a}(t) \hat{\psi}_y^b(t)}
= \sum_{k=0}^2 \ev{\hat{\psi}_x^{-a}(t) \hat{\psi}_y^b(t)}_{k-\text{loc}} .
}
We have already studied the fully extended piece, $\ev{\hat{\psi}_x^{-a}(t) \hat{\psi}_y^b(t)}_{0-\text{loc}}$, in detail in Section~\ref{sec:2pt_relaxation}. It generically relaxes to its GGE value as $t \to \infty$ in a manner described by Eq.~(\ref{eq:gen_dCt}) or (\ref{eq:gen_dCjt}), and this relaxation is due to ``interference'' effects. On the other hand, the $1-$loc piece vanishes as $t \to \infty$ for simpler ``phase space'' reasons: the results of Section~\ref{sec:loc_gaussification}, in particular Eq.~(\ref{eq:1_loc_decay}), show that
\eq{
\ev{\hat{\psi}_x^{-a}(t) \hat{\psi}_y^b(t)}_{1-\text{loc}} \sim [\mathcal{V}(t)]^{-1/2} 
\quad \ \text{as} \quad t \to \infty .
}
This piece is fully off-diagonal in the quasiparticle basis of $\hat{H}$, so its GGE value is also zero. 

Therefore, we only need to study the fully localized ($2-$loc) piece. It is given by (compare Eq.~(\ref{eq:connected_n-loc})):
\eqal{\label{eq:2pt_loc}
\ev{\hat{\psi}_x^{-a}(t) \hat{\psi}_y^b(t)}_{2-\text{loc}} \vphantom{\sum} \qquad\quad \nonumber \\*
= \sum_{j_1, j_2} \, \sum_{c_1, c_2 = \pm} \Big\{
&[R_{x j_1}^{a c_1}]^*R_{y j_2}^{b c_2} \, e^{-i (c_1 \omega_{j_1} - \, c_2 \omega_{j_2}) t} \nonumber \\*[-0.6em]
&\times \ev{\hat{\gamma}^{-c_1}_{j_1} \hat{\gamma}^{c_2}_{j_2}} \Big\} ,
}
where the sums over $j_1$ and $j_2$ are over all levels in the discrete part of the spectrum. Recall (Section~\ref{sec:gge_rho}) that the mode operators $\hat{\gamma}_j$ can (and should) be chosen so that, in each degenerate subspace (i.e.~when $\omega_{j_1} = \omega_{j_2}$), one has $\ev{\hat{\gamma}^{-c_1}_{j_1} \hat{\gamma}^{c_2}_{j_2}} \propto \delta_{c_1 c_2} \delta_{j_1 j_2}$. This ensures that the infinite time-average of Eq.~(\ref{eq:2pt_loc}) agrees with its GGE value:
\eq{
\ev{\hat{\psi}_x^{-a} \hat{\psi}_y^b}^{\text{GGE}}_{2-\text{loc}} = 
\sum_{j} \sum_{c = \pm} \,
[R_{x j}^{a c}]^* \, R_{y j}^{b c} \, \ev{\hat{\gamma}^{-c}_{j} \hat{\gamma}^{c}_{j}} .
}
In general, the instantaneous difference
\eq{
[\delta C_{2-\text{loc}}(t)]_{xy}^{ab} \coloneqq
\ev{\hat{\psi}_x^{-a}(t) \hat{\psi}_y^b(t)}_{2-\text{loc}}
- \ev{\hat{\psi}_x^{-a} \hat{\psi}_y^b}^{\text{GGE}}_{2-\text{loc}} 
\vphantom{\bigg|}
}
oscillates forever about zero without relaxing as $t \to \infty$. However, if the localized states are located far enough apart in space that the initial correlations $\ev{\hat{\gamma}^{-c_1}_{j_1} \hat{\gamma}^{c_2}_{j_2}}$ between them are negligible (this must hold for all nondegenerate pairs $j_1 \neq j_2$), then it follows that $\delta C_{2-\text{loc}}(t)$ is negligible at all times. We conclude that:

Dynamics generated by a quadratic fermion hamiltonian $\hat{H}$ whose quasiparticle spectrum includes discrete localized levels \emph{will still lead to gaussification and equilibration to the GGE}, as long as (i) the initial state has a finite correlation length $\xi$, and (ii) the spatial distance between any pair of localized levels is large relative to $\xi$.

\section{Time-periodic hamiltonians and the ``Floquet-GGE''}\label{sec:floquet_gge}

Our arguments for gaussification in Section~\ref{sec:gaussification} were extremely general; they relied only on clustering of correlations in the initial state, and on spreading of the propagator $G(t)$. Thus, they apply to any quadratic hamiltonian $\hat{H}(t)$, as long as it leads to delocalizing dynamics. In this section, we consider the particularly interesting \emph{time-periodic} case:
\eq{
\hat{H}(t) = \tfrac{1}{2} \hat{\Psi}^{\dagger} \mathcal{H}(t) \hat{\Psi} + \text{constant} ,
}
where
\eq{
\mathcal{H}(t) = \mat{ h(t) & \Delta(t) \\ \pm \Delta^*(t) & \pm h^*(t) } = \mathcal{H}(t+T) ,
}
and where, as before, the plus (minus) sign is for bosons (fermions). $\hat{H}(t)$ describes a periodically driven, or ``Floquet'', closed quantum system.

\subsection{Floquet theory basics}

Let us briefly review some simple facts about this problem \cite{Bukov2015}. In order to give a complete ``stroboscopic'' description of the system at times $t = n T$ ($n = 0,1,2,\cdots$) one only needs to know the time-evolution operator over a single period,
\eq{
\hat{U}(T) = \mathcal{T} e^{-i \int_0^T \hat{H}(t') \dd{t'}}
}
(to describe the system at intermediate times, one also needs to know $\hat{U}(t)$ for all $0 < t < T$). Since $\hat{U}(T)$ is unitary, it has a spectral decomposition of the form
\eq{
\hat{U}(T) = \sum_{\alpha}  e^{-i \epsilon_{\alpha} T} \op{\alpha} ,
}
where $\{ \ket{\alpha} \}$ forms a basis for the Hilbert space of the system, and where the ``quasienergies'' $\epsilon_{\alpha}$ are defined modulo $2\pi/T$. 
The associated ``Floquet hamiltonian''
\eq{
\hat{H}_F \coloneqq \sum_{\alpha} \epsilon_{\alpha} \op{\alpha}
}
generates
\eq{
\hat{U}(T) = e^{-i \hat{H}_F T}
}
by construction. $\hat{H}_F$ is quadratic because $\hat{H}(t)$ is quadratic (quadratic forms in fermion/boson operators form a Lie algebra, and the unitary group is compact). Thus, apart from the subtlety that the quasienergies $\{ \epsilon_{\alpha} \}$ take values on a circle rather than on the real line, the dynamical problem at times $t = nT$ is formally identical to one with a time-independent quadratic hamiltonian
\eq{
\hat{H}_F = \tfrac{1}{2} \hat{\Psi}^{\dagger} \mathcal{H}_F \hat{\Psi} + \text{constant} ,
}
where
\eq{\label{eq:floquet_H_mat}
\mathcal{H}_F = \mat{ h_F & \Delta_F \\ \pm \Delta^*_F & \pm h^*_F } .
}

\subsection{Propagator and gaussification}

Recall that the propagator $G(t)$ is defined by the solution of the Heisenberg equations of motion,
\eq{
\hat{\Psi}(t) = G(t) \hat{\Psi}(0) .
}
Since $\hat{\Psi}(t) \coloneqq \hat{U}^{\dagger}(t) \hat{\Psi}(0) \hat{U}(t)$, and since $\hat{U}(T) = \mathcal{T} e^{-i \int_0^T \hat{H}(t') \dd{t'}} = e^{-i\hat{H}_F T}$, one obtains two equivalent expressions for the propagator over one period:
\eq{
G(T) = \mathcal{T} e^{-i \int_0^xT M(t') \dd{t'}} = e^{-i M_F T} ,
}
where
\eq{
M(t) = \mat{h(t) & \Delta(t) \\ - \Delta^*(t) & -h^*(t)}
}
and
\eq{\label{eq:floquet_G_mat}
M_F = \mat{ h_F & \Delta_F \\ - \Delta^*_F & - h^*_F } .
}
As one might expect for a quadratic system, $G(T)$ completely determines $\hat{H}_F$ (modulo shifting the quasienergies by multiples of $2\pi/T$). 

The propagator at any time $t = nT + t'$, where $0 \leq t' < T$, is given by
\eq{
G(nT + t') = G(t') [G(T)]^n .
}
As $t \to \infty$, the relaxation behavior will be dominated by the $[G(T)]^n$ factor, except possibly in some pathological cases. Therefore, we expect any local connected 3- or higher-point function of the driven system to relax (or fail to relax) with time in exactly the same manner as that of an undriven system with hamiltonian $\hat{H}_F$, up to a multiplicative periodic factor $f(t) = f(t + T)$ coming from the $G(t')$ part of the propagator. The results of Sections~\ref{sec:rates}, \ref{sec:d=1} and \ref{sec:bosons} may thus be applied with only minor modifications.

\subsection{Relaxation to the Floquet-GGE}\label{sec:floquet_gge_relaxation}

Having discussed gaussification in the Floquet context, let us next consider the eventual fate of the effectively gaussian state. Following Section~\ref{sec:gge_rho}, we may construct a GGE density operator $\hat{\rho}_F$ out of the local conserved charges of the Floquet hamiltonian $\hat{H}_F$; the argument of Section~\ref{sec:conserved_quantities}, applied at stroboscopic times $t = nT$, shows that these charges are all quadratic, so that $\hat{\rho}_F$ is indeed gaussian. It is natural to suspect that the system eventually relaxes to a state described by $\hat{\rho}_F$. Note that such a state is a \emph{limit cycle}: for any operator $\hat{\mathcal{O}}$, one has
\eqal{
\Tr(\hat{\mathcal{O}}(t) \hat{\rho}_F) 
&= \Tr( \hat{U}(T) \hat{\mathcal{O}}(t+T) \hat{U}^{\dagger}(T) \hat{\rho}_F) \nonumber \\*
&= \Tr(\hat{\mathcal{O}}(t+T) \hat{\rho}_F)
}
(the second equality follows from the definition of $\hat{\rho}_F$), but in general
\eq{
\text{Tr}(\hat{\mathcal{O}}(t) \hat{\rho}_F) \neq \text{Tr}(\hat{\mathcal{O}}(t') \hat{\rho}_F) .
}
This time-periodic limiting state has been called the ``periodic Gibbs ensemble (PGE)'' or the \emph{Floquet-GGE} \cite{Lazarides2014,Russomanno2016,Moessner2017}.

We can generalize the analysis of Section~\ref{sec:2pt_relaxation} to study relaxation of the gaussified state to the Floquet-GGE. It is sufficient to study this at stroboscopic times $t = nT$. As in Section~\ref{sec:2pt_relaxation}, we must make an additional assumption on the initial state, Eq.~(\ref{eq:gen_homogeneous}), to exclude situations in which the initial profiles of local conserved densities are inhomogeneous on length scales comparable to the system size. Equations (\ref{eq:gen_Gt}--\ref{eq:gen_2pt_structure}) are unchanged, except that $S$ must now be understood as the matrix of the Bogoliubov transformation $\hat{\Psi} = S \, \hat{\Gamma}$ that diagonalizes the Floquet hamiltonian $\hat{H}_F$, and $\varepsilon$ as a quasienergy defined modulo $2\pi/T$.

First consider the limit of very fast driving, $T \to 0$. In this limit, we expect that we can ignore the periodicity of $\varepsilon$ (since the period $2\pi/T \to \infty$), and that the quasiparticle states of $\hat{H}_F$ are organized into one or more well-defined bands. If this is so, the remainder of the analysis in Section~\ref{sec:2pt_relaxation} applies, and we conclude that the gaussified state relaxes to $\hat{\rho}_F$ with a power law; in the simplest cases, this power law is given by Eq.~(\ref{eq:gen_dCt}) or (\ref{eq:gen_dCjt}).

Next consider the opposite limit of very slow driving, so that $T \to \infty$ and $\hat{H}(t)$ is a slowly varying function of $t$. On timescales $t \lesssim T$, we expect, based on an adiabatic approximation and our arguments in the time-independent case, to observe power-law relaxation to a GGE of the instantaneous hamiltonian $\hat{H}(t)$. On much longer timescales $t \gg T$, the Floquet drive becomes important, and we except to eventually observe relaxation to the Floquet GGE, $\hat{\rho}_F$. Stationary phase analysis suggests that this relaxation will be \emph{exponential} in time, $\sim e^{-t/T}$. To see this, consider Eq.~(\ref{eq:gen_dCt_int}). The quasienergies $\varepsilon$ are defined on a circle of radius $2\pi/T \to 0$, so the spectrum is likely to be relatively smooth, without well-defined bands. Therefore
\eqal{
\delta C(t) &\sim \int_0^{2\pi/T} \!\! \dd{\varepsilon} \int_0^{2\pi/T} \!\! \dd{\varepsilon'}
a(\varepsilon, \varepsilon') \, e^{-i(\varepsilon - \varepsilon')t} \nonumber \\*
&= \frac{1}{T^2} \int_0^{2\pi} \! \dd{z} \int_0^{2\pi} \!\! \dd{z'}
a\Big(\frac{z}{T}, \frac{z'}{T}\Big) e^{-i(z - z')t/T} ,
}
where $a(z/T, z'/T)$ is a smooth function of $z$ and $z'$ on the torus $\mathbb{T}$. It follows that $\delta C(t)$ must vanish faster than any power of $(t/T)$ as $(t/T) \to \infty$. Similar arguments apply to $\delta C_j(t)$.

Thus, in the limit $T \to 0$ of fast driving, we expect to observe power-law relaxation to $\hat{\rho}_F$, the (time-periodic) GGE of the Floquet hamiltonian $\hat{H}_F$. In the opposite limit $T \to \infty$ of slow driving, we expect to observe power-law relaxation toward a GGE of the instantaneous hamiltonian $\hat{H}(t)$, followed by much slower exponential relaxation $\sim e^{-t/T}$ toward $\hat{\rho}_F$. It is more difficult to make semi-quantitive general statements about the regime of intermediate driving, and we leave this as an interesting question for future work.

\section{A comment on spin models mappable to quadratic fermion models}\label{sec:spins}

Everything that we have said also applies to any spin system that can be mapped to a quadratic model of fermions (via a Jordan-Wigner transformation or otherwise), assuming (i) that the observables of interest map to local operators in terms of the fermions, and (ii) that the initial state $\hat{\rho}_0$ obeys cluster decomposition \emph{with respect to the fermion operators}. It is by no means obvious that a given physical initial state, which obeys cluster decomposition with respect to spin operators, also does so with respect to the fermions. It would be interesting to identify which states have this property.

\section{Conclusions}\label{sec:conclusions}

We have presented a general framework for understanding relaxation phenomena in systems described by quadratic fermion or boson hamiltonians that may or may not be time-dependent. We have shown that, as long as the hamiltonian yields delocalizing dynamics, and for any initial state that satisfies a condition on algebraic clustering of correlations, all local operators of the system relax to values consistent with a gaussian state at late times---the system ``gaussifies''. Furthermore, we have shown that gaussification can be understood as a simple consequence of the spreading of operators in real space, and that the exponents of the power laws with which quantities gaussify can be extracted from the smooth envelope of the one-particle propagator of the system (which does not depend on the initial state). In this sense, gaussification in quadratic systems appears to be quite universal in character.

Using similar arguments, we have given a simple proof that all local conserved quantities of a quadratic time-independent hamiltonian with delocalizing dynamics are themselves quadratic, and hence that the GGE density operator of such a system is gaussian. We have described how to construct the GGE out of mode occupation numbers in a manner that properly accounts for degeneracies in the mode spectrum. Under an additional assumption on the initial state (needed to avoid having to deal with hydrodynamic timescales comparable to the system size), we have shown that the local 2-point function of the system relaxes to its GGE value with a power law whose exponent can typically be extracted from the local density of single-particle levels at the band edge. Combined with our gaussification results, this proves relaxation to the GGE for a large class of quadratic systems and a large family of initial states, and also gives quantitative information about how local observables relax. We find that, if the initial state has a density wave of some conserved quantity, the system generically relaxes first to a gaussian state, and then, with a smaller inverse power of time, to the GGE. If the initial state is \emph{not} ordered in this sense, ``gaussification'' and relaxation to the GGE occur with the same powers of time and cannot be distinguished as easily in general.

We have also studied situations in which these conclusions break down, such as the case of free massless bosons in one dimension \cite{Sotiriadis2016,Sotiriadis2017}, or when the mode spectrum of the hamiltonian includes localized levels, and have explained precisely why the breakdown occurs in these cases. We have argued that, perhaps unexpectedly, well-separated localized levels in a system of fermions do not hinder gaussification or relaxation to the GGE. Finally, we have applied our arguments to the case of periodically driven systems, and have shown that the relaxation of such systems to the Floquet-GGE can also be understood semi-quantitatively within our framework.

\section*{Note added}

Shortly after a preprint of this work was first posted, another preprint presenting closely related and complementary results appeared \cite{Gluza2018}.
Our work places more emphasis on physical intuition, and on estimating in a simple manner the exponents of the power laws by which local observables relax in quadratic models, while Ref.~\cite{Gluza2018} places more emphasis on rigorous results and error bounds.
The methods used are also different (though related in spirit): where we employ the stationary phase approximation, Ref.~\cite{Gluza2018} uses the machinery of Kusmin-Landau bounds.

\begin{acknowledgments}

We thank Fabian H.~L.~Essler, Matthew P.~A.~Fisher, Tarun Grover, Chetan Nayak, Shivaji Sondhi, and especially Marcos Rigol for helpful discussions. This work was supported in part by NSF Grant PHY13-16748, and in part by the Microsoft Corporation Station Q (C.M.).

\end{acknowledgments}


\bibliography{gge_refs}


\appendix

\section{Connected correlation functions}\label{app:cumulants}

For completeness, in this section we review the standard definition of a connected correlation function \cite{Weinberg1995}.

Let $\ev{\hat{X}} \coloneqq \text{Tr}(\hat{X} \hat{\rho})$ denote the expectation of the operator $\hat{X}$ in a given state $\hat{\rho}$. The \emph{connected correlation function} or \emph{cumulant} $\cev{\cdots}$ of a set of operators $\hat{X}_1, \hat{X}_2, \dots, \hat{X}_n$ is defined inductively by the formula
\eq{
\ev{\hat{X}_1 \hat{X}_2 \cdots \hat{X}_n} = \sum_{P} (\pm) \prod_{\alpha \in P} \cev{\hat{X}_{\alpha(1)} \hat{X}_{\alpha(2)} \cdots } ,
}
where the sum is over all partitions $P$ of the set $\{1,2,\dots,n\}$, each element $\alpha_j$ of the partition is ordered so that $\alpha_j(1) < \alpha_j(2) < \cdots $, and the sign is $+$ or $-$ according to whether the rearrangement
\eqml{
(1,2,\cdots,n) \mapsto \\*
(\alpha_1(1),\alpha_1(2), \cdots, \alpha_2(1), \alpha_2(2), \cdots, \cdots \cdots)
}
involves altogether an even or odd number of exchanges of fermionic operators, respectively. Unpacking the definition for small values of $n$,
\eqsub{
\label{cumulant_def_123}
\eqal{
\ev{\hat{X}_1} = \ 
&\cev{\hat{X}_1} , \\*
\ev{\hat{X}_1 \hat{X}_2} = \ 
&\cev{\hat{X}_1 \hat{X}_2} + \cev{\hat{X}_1} \cev{\hat{X}_2} , \\*
\ev{\hat{X}_1 \hat{X}_2 \hat{X}_3} = \ 
&\cev{\hat{X}_1 \hat{X}_2 \hat{X}_3} 
+ \cev{\hat{X}_1 \hat{X}_2} \cev{\hat{X}_3} \nonumber \\*
&+ \cev{\hat{X}_1} \cev{\hat{X}_2 \hat{X}_3} 
\pm \cev{\hat{X}_1 \hat{X}_3} \cev{\hat{X}_2} \nonumber \\*
&+ \cev{\hat{X}_1} \cev{\hat{X}_2} \cev{\hat{X}_3} ,
}
}
and so on.

Informally, the cumulant $\cev{\hat{X}_1 \hat{X}_2 \cdots \hat{X}_n}$ equals the correlation function $\ev{\hat{X}_1 \hat{X}_2 \cdots \hat{X}_n}$, minus all possible ways of factorizing this function into products of two or more lower-order cumulants (with additional minus signs as needed to account for exchanges of fermionic operators).

\section{Bounds on $g(t)$ for bosons with pairing}\label{app:bounds}

Any quadratic time-independent hamiltonian $\hat{H} = \frac{1}{2} \hat{\Psi}^{\dagger} \mathcal{H} \hat{\Psi}$ for a system of bosons, in which $\mathcal{H}$ is positive-definite, can be diagonalized by a Bogolyubov transformation \cite{VanHemmen1980}:
\eq{
\hat{\Psi} = S \, \hat{\Gamma} ,
}
where
\eq{
\hat{\Gamma} = (\hat{\gamma}^-_1, \hat{\gamma}^-_2, \cdots, \hat{\gamma}^-_N, \hat{\gamma}^+_1, \hat{\gamma}^+_2, \cdots, \hat{\gamma}^+_N)^T .
}
The transformation $S$ has the block form
\eq{
S = \mat{U & V^* \\ V & U^*} ,
}
satisfies $S^{\dagger} \eta S = \eta$, where $\eta = I_N \oplus - I_N$, and diagonalizes the hamiltonian matrix: $S^{\dagger} \mathcal{H} S = \eta \Omega$, where
\eq{
\Omega \coloneqq \text{diag}(\omega_1, \omega_2, \dots, \omega_N, -\omega_1, -\omega_2, \dots, -\omega_N) .
}
Since $\mathcal{H}$ is positive-definite, all $\omega_j > 0$ (by Sylvester's theorem of inertia). The condition $S^{\dagger} \eta S = \eta$ may be rewritten as $S^{\dagger} = \eta S^{-1} \eta$. It follows that
\eq{
\eta \mathcal{H} = S \Omega S^{-1} .
}
In terms of the quasiparticle operators, we have
\eq{
\hat{H} = E_0 + \sum_{j=1}^N \omega_j \hat{n}_j ,
}
where $\hat{n}_j = \hat{\gamma}^+_j \hat{\gamma}^-_j$. 

It is very important to note that, since $S$ is not in general a unitary transformation, the boson mode energies $\omega_j$ are \emph{not} the eigenvalues of the hermitian matrix $\mathcal{H}$. We will denote the eigenvalues of the matrix $\mathcal{H}$ as $\epsilon_j$. We recover $\omega_j = \epsilon_j$ only when all pairing terms in the hamiltonian vanish; in this limit $S$ is indeed unitary.

The propagator $G(t)$ may be written in matrix form as
\eqal{
G(t) &= e^{-i \eta \mathcal{H} t} \nonumber \\*
&= S e^{-i \Omega t} \, S^{-1} \nonumber \\*
&= S e^{-i \Omega t} \eta S^{\dagger} \eta .
}
At each time $t$, it satisfies $G(t) \eta \, G^{\dagger}(t) = \eta$. From this fact, we easily obtain the lower bound $1 \leq g^a_x(t)$, as follows (no sum on $x,a$):
\eqal{\label{eq:boson_gt_lower_bound}
1 &= \abs{\eta^{aa}_{xx}} \nonumber \\*
&= \abs{ \sum_{b = \pm} \sum_y [G(t)]^{ab}_{xy} \, \eta^{bb}_{yy} \, [G^{\dagger}(t)]^{ba}_{yx} } \nonumber \\*
&\leq \sum_{b = \pm} \sum_y \abs*{G^{ab}_{xy}(t)}^2 \nonumber \\*
&= g^a_x(t) .
}

More work is required to derive an upper bound on $g^a_x(t)$. We have
\eqal{
g^a_x(t) &= \sum_{b=\pm} \sum_{y} \abs*{G^{ab}_{xy}(t)}^2 \nonumber \\*
&= [G^{\dagger}(t) G(t)]^{aa}_{xx} \qquad \text{(no sum on $x,a$)} .
}
By definition of the operator norm $\norm{\cdot}$,
\eq{
g^a_x(t) \leq \norm*{G^{\dagger}(t) G(t)} .
}
For bounded operators $A$ and $B$, one has $\norm{AB} \leq \norm{A} \norm{B}$ and $\norm*{A^{\dagger}} = \norm{A}$. Since $\norm{\eta} = \norm*{e^{- i \Omega t}} = 1$, it follows that
\eq{
g^a_x(t) \leq \norm{S}^4 .
}

We can derive a bound on $\norm{S}$ from the condition $S^{\dagger} \mathcal{H} S = \eta \Omega = \text{diag}(\omega_1, \omega_2, \dots, \omega_N, \omega_1, \omega_2, \dots, \omega_N)$. Since the hermitian matrix $\mathcal{H}$ is positive-definite, it has a unique positive-definite square root, $\mathcal{H}^{1/2}$. Let $R \coloneqq \mathcal{H}^{1/2} S$, so that $R^{\dagger} R = \eta \Omega$. The operator norm of $R$ equals the square root of the largest eigenvalue of $R^{\dagger} R$, so
\eq{
\norm{R}^2 = \omega_{\text{max}} \coloneqq \max \{ \omega_1, \omega_2, \cdots, \omega_N \} .
}
Since $S = (\mathcal{H}^{1/2})^{-1} R$, it follows that
\eq{
\norm{S} \leq \norm*{(\mathcal{H}^{1/2})^{-1}} \norm{R} .
}
The operator norm of $(\mathcal{H}^{1/2})^{-1}$ equals the square root of the largest eigenvalue of $((\mathcal{H}^{1/2})^{-1})^{\dagger} (\mathcal{H}^{1/2})^{-1} = \mathcal{H}^{-1}$, so 
\eq{
\norm*{(\mathcal{H}^{1/2})^{-1}} = \frac{1}{\sqrt{\epsilon_{\text{min}}}} ,
}
where $\epsilon_{\text{min}} = \min \{ \epsilon_1, \epsilon_2, \cdots, \epsilon_N \}$ is the smallest eigenvalue of $\mathcal{H}$. Thus,
\eq{
\norm{S} \leq \sqrt{\frac{\omega_{\text{max}}}{\epsilon_{\text{min}}}} ,
}
and we finally obtain the desired upper bound:
\eq{\label{eq:boson_gt_upper_bound}
g^a_x(t) \leq \left( \frac{\omega_{\text{max}}}{\epsilon_{\text{min}}} \right)^2 .
}
Equations~(\ref{eq:boson_gt_lower_bound}) and (\ref{eq:boson_gt_upper_bound}) together yield Eq.~(\ref{eq:gt_bound_bosons}).

\end{document}